# Prominent but Less Productive: The Impact of Interdisciplinarity on Scientists' Research

Erin Leahey
School of Sociology
University of Arizona
Tucson, AZ 85721
leahey@email.arizona.edu

Christine M. Beckman
School of Business
UC Irvine and University of Maryland
College Park, MD
(301) 405-9002
cbeckman@umd.edu

Taryn L. Stanko
Orfalea College of Business
California State Polytechnic University
San Luis Obispo, CA 93407
tstanko@calpoly.edu

DRAFT
30 July 2015

The first two authors listed are equal coauthors. This research was supported by NSF award #0332051 in 2003. We thank the seminar participants at University of Chicago, Penn State, McGill, Carnegie Mellon, University of Maryland, UC Irvine, UC Riverside, and University of Arizona for their helpful comments. Earlier versions of this paper were presented at the American Sociological Association in 2012, and the Academy of Management Meetings in 2014.

**Prominent but Less Productive: The Impact of Interdisciplinarity
on Scientists' Research**

ABSTRACT

Inter-disciplinary research (IDR) is being promoted by federal agencies and universities nationwide because it presumably spurs transformative, innovative science. In this paper we bring empirical data to assess whether IDR is indeed beneficial, and whether costs accompany potential benefits. Existing research highlights this tension: whereas the innovation literature suggests that spanning disciplines is beneficial because it allows scientists to see connections across fields, the categories literature suggests that spanning disciplines is penalized, because the resulting research may be lower quality or confusing to place. To investigate this, we empirically distinguish production and reception effects and we highlight a new production penalty: cognitive and collaborative challenges associated with IDR may result in slower progress, hurdles during peer review, and lower productivity (though not necessarily lower quality). We compile and analyze data on almost 900 research center-based scientists and their 32,000 published articles. Using an innovative measure of IDR that considers the similarity of the disciplines spanned, we document both penalties (fewer papers published) and benefits (increased visibility) associated with IDR, and show that it is a high-risk, high-reward endeavor. These costs and benefits depend on characteristics of the field and a scientist's place in it.

### Prominent but Less Productive:
### The Impact of Interdisciplinarity on Scientists' Research

Because of its expected benefits to science and society (Rhoten and Parker 2004; Sanz, Bordons, and Zulueta 2001), scholars increasingly engage in an interdisciplinary mode of research, which "integrates perspectives, information, data, techniques, tools, concepts, and/or theories from two or more disciplines" (National Academies of Science, National Academy of Engineering, and Institute of Medicine 2005:188). While the practice of interdisciplinarity is not new (Abbott 2001), it is increasingly prevalent in the natural (Rhoten and Pfirman 2007) and social sciences (Brint 2005; Jacobs and Frickel 2009). Universities are reorganizing to facilitate **interdisciplinary research (IDR)** by developing cross-disciplinary problem-focused centers and funding cross-department and cross-college research initiatives (Biancani, McFarland, and Dahlander in press; Pray 2002). And since the mid-1980s, the National Science Foundation has supported cross-cutting funding opportunities and interdisciplinary research centers. Scientists laud IDR as a 'progressive' 'hot topic' that is 'running rampant;' arguably one 'must be interdisciplinary to be world-class' (Pray 2002). But evidence in support of this contention is sparse and "relatively little research on many of the underlying issues has been conducted" (Jacobs and Frickel 2009: 44). In particular, systematic investigation of IDR's effects on scientific careers has been neglected. What are the professional costs and benefits of engaging in IDR?

To understand the impact of IDR on scientists' careers, we draw on two strands of organizational theory. The first documents the innovative <u>benefits</u> of joining diverse ideas across domains. The core idea can be found in theories of diversity, brokerage, and innovation (Burt 2004; Singh and Fleming 2010): pooling non-redundant information from disparate sources is the foundation from which novel ideas spring (Hargadon 2002; Weitzman 1998). This work suggests that bridging disconnected knowledge spaces will result in better ideas that will be rewarded in the marketplace (Lo and Kennedy in press). The second documents <u>penalties</u> associated with category spanning (Zuckerman 1999). Rather than being perceived as innovative, offerings spanning multiple domains have an ambiguous identity that is difficult for audiences to understand, and are thus devalued (Hsu, Hannan, and Kocak 2009). Recent empirical work describes the negative perceptions of the audience or market as *reception-side* penalties. However, the category literature has also theorized *production-side* penalties where category-spanning products are more difficult to produce. The principle of allocation suggests that investing in multiple categories limits mastery and dilutes quality, resulting in a "Jack of all trades" who is master of none (Hannan and Freeman 1989; Hsu 2006b). This work builds on niche-width theory that anticipates problems of focus when organizations span strategies or forms (Freeman and Hannan 1983). Recent empirical work has found that category-spanning products are lower quality, likely because the production process itself is wanting (Kovács and Johnson in press). This contributes to, but does not completely explain, the reception-side devaluation of category-spanning products (Kovacs and Sharkey 2014; Negro and Leung 2013). From this we gather that production- and reception-side effects can operate simultaneously.

On the production side, we identify limitations in previous scholarship on penalties that accrue to category-spanning offerings. We suggest production penalties result not only from limited mastery and inferior quality, which are highlighted in the categories literature (Hsu, Hannan, and Kocak 2009; Negro and Leung 2013), but also from the cognitive and collaborative challenges of category-spanning work (Rafols, Leydesdorff, OHare, Nightingale, and Stirling 2012). Grasping ideas and perspectives from another field is cognitively taxing, working with diverse collaborators from multiple disciplines can produce frustration and conflict, and reviewers may have difficulty digesting and evaluating cross-disciplinary products. All of these challenges may lengthen the time to publication, and thereby depress scholars' productivity.

On the reception-side, we expect work that draws on disparate intellectual domains to have broad appeal and achieve greater scholarly visibility. Atypical and novel combinations of ideas have greater impact (Schilling and Green 2011), so scientists may be drawn toward, rather than confused by, multi-category offerings like IDR, like Pontikes' (2013) market-makers. Following the innovation literature which finds that experimenting with new combinations is uncertain and risky (Fleming 2001), we similarly expect that IDR, like other high-risk strategies, sustains overall benefits but higher variance in reception. Although IDR boosts visibility overall (mean levels of citations), it should also increase the *variance* of citations that a scientist's body of work garners.

Further, we refine the way in which category-spanning is typically measured. Rather than spanning two categories (i.e., crossing a single boundary), interdisciplinary research typically draws upon multiple disciplines. Theoretically, we begin with disciplinary *variety* as the underlying concept of interest to us (Harrison and Klein 2007). Methodologically, we operationalize this not by assessing whether a line has been crossed, as others have done (Fleming, Mingo, and Chen 2007; Hsu, Hannan, and Kocak 2009), but by measuring the variety inherent in each scientist's body of work. While this is an advance over binary measures of IDR (Clemens, Powell, McIlwaine, and Okamoto 1995; Jacobs and Frickel 2009), we take it one step further – following Leahey & Moody (2014) – to account for the cognitive distance between the categories (i.e., disciplines) because all disciplinary combinations are not equally disparate. In this way, we build on recent work in the category literature that has begun to examine the similarity of (and distance between) categories (Hannan and Kovacs 2011; Kovacs and Sharkey 2014).

Finally, we suggest that the penalties and benefits of interdisciplinary work depend upon the nature and intellectual life cycle of one's field. The difficulties associated with category-spanning may be reduced, and the benefits may be accentuated, when category-spanning is popular (Lo and Kennedy in press). In fields with a tradition of IDR, training may reduce the difficulties of producing IDR research and audiences may particularly value such work. Thus we assess not only the main effects of IDR on productivity and visibility, but also how field-level norms modify such effects.

In sum, this study makes several contributions to organizational theory. First, we empirically distinguish between – and assess the simultaneous effects of – production side and reception side processes. Second, we highlight production penalties that result from cognitive and collaborative challenges (as Zuckerman and colleagues (2003) anticipated) and disentangle potential mechanisms that manifest in different stages of the production process. Third, we move away from a binary conceptualization of 'spanning' and toward a continuous conceptualization of 'variety' that uniquely considers the dissimilarity among categories (here, disciplines). Fourth, we begin to understand the normalization of category-spanning by examining an important, contextual-level moderator of such effects: field-level interdisciplinarity. We further explicate the nature of these contributions, and then proceed to test our ideas using primary data collected on almost 900 scientists and their 30,000 scholarly papers.

PRODUCTION PENALTIES OF INTERDISCIPLINARY RESEARCH

Disciplinary fields are an entrenched way in which scholarly activity is organized (Sá 2008) and evaluated (Lamont 2009), making field-spanning offerings unexpected. Like other categories, fields are not arbitrarily constructed, but rather reflect the distinct environments that disciplines face, the formal and tacit skill sets that members acquire (Zuckerman, Kim, Ukanwa, and Rittmann 2003), and the distinct claims of jurisdiction that occasionally come into dispute (Abbott 2001). Disciplinary fields are categories that help academics parse and digest a vast intellectual terrain in order to conduct and evaluate scholarship. From the scholarship on categories, we know that when categories are not strictly adhered to, penalties ensue. Theoretically, penalties can emerge from both the production side and the reception side. For example, Zuckerman, Kim, Ukanwa, and Rittmann (2003), suggest that penalties emanate from both the production side (e.g., it's harder to produce and quality may suffer) and the reception side (e.g., it is confusing for audiences to place a category-spanning offering). Recent empirical tests have focused mostly on the reception penalties: audiences tend to overlook, devalue, or outright reject offerings that span categories because they are difficult to comprehend and do not fit existing schemas (Hsu, Hannan, and Kocak 2009). Category-spanning actors and offerings tend to be poorly received: feature-film actors who take on too many diverse roles have difficulty obtaining work (Zuckerman, Kim, Ukanwa, and Rittmann 2003); films spanning multiple genres are less appealing to audiences (Hsu 2006a); and eBay sellers who try to market their product in multiple categories are less successful in the auction (Hsu, Hannan, and Kocak 2009). Audiences penalize offerings that are difficult to classify.

We turn our attention to less examined production penalties and in particular to the cognitive and collaborative challenges that slow the development of mastery. Because cognitive resources are finite, the argument goes, mastery may be more difficult to attain when time, energy, and effort are distributed across many categories (Carroll 1985; Freeman and Hannan 1983; Hannan and Freeman 1989; Hannan, Polos, and Carroll 2007). This early work in population ecology theorized, but did not empirically examine, the problems of diffused resources and focus that may disadvantage generalists in production – including greater competition and likelihood of mortality. Recent work does not examine mastery per se, but suggests that it accounts for lower quality products (Kovacs and Sharkey 2014; Negro and Leung 2013). We hew closer to the concept of mastery itself and argue for technical and logistical hurdles when spanning domains (Lingo and O'Mahony 2010; Zuckerman 2005). For example, research on patents suggests that category-spanning increases technological uncertainty and creates cognitive limitations (Fleming 2001; Lo and Kennedy in press). Lamont and colleagues (2006)lend support to the idea that category-spanning ideas are harder to produce: it is challenging for scholars to accommodate the concepts of multiple fields and to produce output that is standard in form and content (see also Rafols et al. 2012).

In addition, when collaborative in nature, interdisciplinary research may produce epistemological or methodological conflict (Murray 2010; van Rijnsoever and Hessels 2011) between members of different fields. This lack of consensus may increase coordination costs (Cummings and Kiesler 2005; Cummings and Kiesler 2007; Shrum, Genuth, and Chompalov 2007), conflict and tension (Murray 2010; Owen-Smith and Powell 2003), and role strain (Boardman and Bozeman 2007) that may outweigh the general benefits of collaborative work (Gans and Murray 2014). Because commonality and consensus speed the implementation of ideas (Beckman 2006), such interdisciplinary divergences and tensions may slow progress toward publication.

These initial production penalties, experienced as scientists plan, conduct, and write up their research, likely follow the paper into a second stage of production: the review stage. In the peer review process, experts from different fields often disagree on the merits of a paper and evaluate them differently (Lamont 2009; Mansilla 2006). Indeed, Birnbaum (1981) finds that research that does not fit neatly within the substantive bounds of 'normal science' instills irritation, confusion, and misunderstanding among reviewers and editors. This makes the road to publication challenging, and the final product may take longer to produce. We see a similar phenomenon in the patent context: approval times are longer when inventions span (or in their language, blend) categories (Lo and Kennedy in press). Thus the review process may be slower and more challenging for interdisciplinary work, which contributes to lower rates of productivity.[1]

Indeed, even among highly successful interdisciplinary scholars, we find some initial evidence that production penalties result from cognitive challenges. In concurrent interviews the first author is conducting with recipients of the Mellon Foundation's New Directions fellowship (who, post-tenure in one field, received formal training in another field), we find evidence that interdisciplinary research takes additional time, commitment, and effort. It's mentally more taxing: "I'll be reading epistemology and political philosophy….and then do research and mathematical things. It's just a bit of a stretch of the brain to do such things."  It takes longer: "I think part of me taking longer to go from conception to a finished published article has to do with trying to think through two separate disciplinary concerns."  In the end, it can take a toll on one's productivity: "I just don't feel like I've made great strides of improvement in being as productive as I could be." This suggests that scientists who are more engaged with interdisciplinary work will publish fewer research articles than scientists who engage in IDR less. Thus, we hypothesize a production penalty:

> *H1:  IDR is associated with lower productivity*

RECEPTION BENEFITS OF INTERDISCIPLINARY RESEARCH

Although category-spanning products are typically discounted or dismissed, we do not expect IDR to incur a reception penalty – that is, to be received poorly in the academic marketplace (i.e., rarely cited). As Pontikes (2012) demonstrates, research documenting penalties has largely been conducted on 'market-takers' – audiences that are looking to identify, place and evaluate a product. To market-takers, multiple categories (and ambiguous classification more generally) may produce confusion and prevent a timely and effective search. In contrast, 'market-makers' are looking to identify novel, original, and groundbreaking research that redefines the (academic) market structure (Pontikes 2012). This distinction helps reconcile the reception penalties noted in the categories literature with the reception benefits noted in the recombinatory innovation literature, like Fleming's research documenting that patents that span patent classes and sub-classes receive more citations (Fleming 2001; Singh and Fleming 2010). Market-makers see ambiguous classification not as confusing but rather appealing, and perhaps an indication of innovative work that has brought insights from one arena to illuminate another.

---

[1] We collect data on a sample of unpublished papers to assess whether IDR is more likely to be rejected outright. We also bring additional data on time in the review process to bear on this question.

In fact, various streams of research document the *benefits* that accrue to joining domains of knowledge. For example, Burt (2004) found that when managers span multiple departments, their ideas were evaluated more favorably than managers who didn't span departments. Similarly, research on innovation suggests that information pooled from disparate sources provides a (if not *the*) foundation from which new ideas spring (Fleming and Waguespack 2007; Hargadon 2002). This link is so tight as to warrant use of the term "recombinant innovation" (Weitzman 1998). The underlying idea is that bridging knowledge domains and developing new combinations serves as the foundation for innovation (Schumpeter 1912 [1934]; Weick 1979) and will be recognized as such. For example, Singh and Fleming (2010) find that experience diversity and network size of a research team predict breakthrough patents. Research on the impact of category-spanning in science finds that atypical, category-spanning offerings have higher impact (Lo and Kennedy in press; Schilling and Green 2011; Shi, Adamic, Tseng, and Clarkson 2009; Uzzi, Mukherjee, Stringer, and Jones 2013). Interdisciplinary publications, as a form of atypical, domain-spanning publications, likely experience these same benefits.

In academe, positive reception is typically gauged via citations (Evans 2008; Lynn 2014; Merton 1977). Indeed, citations give us a sense of how a paper is received by the scientific community. Among other things,[2] reference to a scholarly work indicates its usefulness and influence because it contributed, in some way, to a subsequent work. A scientist's citation count gives a good indication of his or her *visibility* in the scientific community, and for this reason it is factored into promotion and tenure decisions, and has been shown to impact earnings (Diamond 1986; Leahey 2007; Sauer 1988). One goal of this paper is to empirically test for such a reception benefit: whether scholarly visibility (as indicated by citation counts) accrues to interdisciplinary work, even after controlling for the size of the prospective audience (which increases when multiple audiences are targeted).

*H2: IDR is associated with higher visibility*

Because we expect to find empirical support for both a production penalty (H1) and a reception benefit (H2), it is likely that IDR is a high-risk, high-reward activity that often gains prominence but sometimes does not. Even if IDR positively impacts visibility (H2), it likely increases the probability of both breakthroughs (like achieving scientific prominence) and failures (like being ignored); (Singh and Fleming 2010) document a similar effect among patents. Thus, although on average IDR will be well-received, we also expect IDR work to have greater variability in citations.

*H3: IDR is associated with greater variability in visibility*

Finally, we highlight that we are distinguishing between spanning categories and spanning *distant* categories. Most investigations simply assess *whether* boundary-spanning has occurred (Clemens, Powell, McIlwaine, and Okamoto 1995; Fleming, Mingo, and Chen 2007; Hsu, Hannan, and Kocak 2009), without considering the relationship between the spanned entities (exceptions include Braun and Schubert (2003), Rosenkopf and Almeida (2003), and Schilling and Green (2011)). The category literature is just beginning to consider distances between categories (Kovacs and Hannan 2011; Kovacs and Sharkey 2014; Leahey and Moody 2014). Are the entities cognitively similar, like civil and chemical engineering? Or are they cognitively distant, like geography and optics? Put simply, interdisciplinary research can be more or less novel, depending on the relationship between the spanned fields (Carnabuci and Bruggerman 2009: 608). Indeed, spanning two closely related categories is hardly different from not spanning at all. In terms of penalties, the cognitive and coordination challenges associated with category-spanning likely increase as the distance between domains increases. In terms of benefits, the utility and value of category-spanning research also increases with distance: the most fertile creative products are "drawn from domains that are far apart" (Poincare 1952 [1908]) and the best conceptual metaphors are those that create ties across great distances (Knorr Cetina 1980). We thus hypothesize that distance matters:

> *H4: An alternate measure of IDR that does not incorporate distance will have weaker effects on productivity and visibility.*

---

[2] There is variation in what a citation signals (van Dalen and Henkens 2005). Citations may reflect disciplinary alliances and mutually-reinforcing citation practices, instrumental attempts to flatter potential reviewers, or even the controversial nature of an article. Authors may cite previous work in a casual way, or rely on it heavily. They may think of it highly, or dismiss it as flawed (Ferber 1986; Latour 1987; Lynn 2014; Najman and Hewitt 2003).

## INTELLECTUAL CONTEXT and IDR

We expect the penalties and benefits associated with IDR to be shaped by each scientist's intellectual context, that is, the interdisciplinary nature of their primary field. We know from previous research that some fields, like the life sciences, are more interdisciplinary than other fields, such as electrical engineering. Data from the Survey of Earned Doctorates (Millar and Dillman 2012) reveals that the life sciences have the highest proportion of dissertations self-classified as interdisciplinary (36%), especially compared to engineering and math (26% and 21% respectively). Using a continuous measure of IDR that incorporates cognitive dissimilarity (distance), Porter and Rafols (2009) and Porter et al. (2007) find that Biotechnology (mean=0.654, on a scale of 0 to 1) and Medicine (mean=0.664) are more interdisciplinary than Electrical Engineering (mean=0.53). Such field-level characteristics and perhaps expectations should influence how a piece of scholarship is received. Actors tend to value what they are accustomed to and what is familiar (McPherson, Smith-Lovin, and Cook 2001; Phillips 2011; Trapido in press). On the reception-side, this suggests that highly interdisciplinary fields, like life sciences, will expect and appreciate IDR work, whereas less interdisciplinary fields will view it as unnecessary or unusual. Thus, scholars in interdisciplinary fields should experience stronger visibility benefits for their IDR work. On the production-side, scholars in highly interdisciplinary fields may experience fewer challenges, as there are models for guidance, potentially better training in IDR, and a more amenable and productive review process for IDR, whereas scholars in less interdisciplinary fields may experience greater difficulties producing IDR. Lo and Kennedy (in press) suggest a similar logic: audience members and patent examiners are more favorable when category-spanning is more familiar and typical. To summarize: if one's disciplinary environment is conducive to (and appreciative of) IDR, then the benefits of IDR might be stronger, and the penalties less stiff:

> *H5a: In highly interdisciplinary fields, IDR's positive association with visibility will be stronger than in less interdisciplinary fields.*

> *H5b: In highly interdisciplinary fields, IDR's negative association with productivity will be weaker than in less interdisciplinary fields.*

In sum, our paper assesses benefits and penalties associated with interdisciplinary research as well as field level expectations that may moderate its effects. We expect the penalties to manifest largely in the production stage, mostly in the form of increased cognitive and communicative challenges, and reduced productivity. Consistent with the innovation literature, we expect benefits to emerge in the reception stage, in the form of increased citations. Our empirical focus heeds Jacobs and Frickel's (2009: 61) call to understand how "interdisciplinary scholarship compare[s] with otherwise similar scholarship" and provides much needed grounding for claims about the value of IDR.

DATA and METHODS

*Sample*

To test our hypotheses, we collected archival publication data on scientists and social scientists associated with 52 NSF-funded industry/university-cooperative research centers (IUCRCs). These centers are housed at universities and conduct research that is of interest to (and partly supported by) industry, in areas as diverse as biosurface physics, civil and environmental engineering, and architectural science. We collected information about all IUCRC members – including doctoral students, post-docs, and faculty at all career stages – as of 2003. Our analysis is limited to the subset of 854 PhD-level scientists with a publication record as of 2005 because our key variables (IDR, productivity, and visibility) cannot be computed for scientists who have not published. Information on approximately 32,000 articles published by these center affiliates (such as number of authors, institutions represented, journal of publication, number of citations) was obtained through Thomson Reuters **Web of Science** (**WoS**). We obtained data on all the articles *referenced* by these 32,000 focal articles, notably their WoS **subject categories (SCs)**, which we used to measure the extent to which each scholar's research is interdisciplinary (described below). To these data we added field-, university-, and individual-level data from various sources (also detailed below). Our main analyses involve these archival data; however, for supplementary analyses, we relied on a survey of these center-based scientists (conducted by the second author), publicly available data on journal turn-around times, and unpublished working papers for a subset of authors.

*Dependent Variables*

**Productivity**. To capture each scholar's productivity, we rely on the total number of articles published (in WoS journals) from the beginning of a scholar's career (i.e., the year they first published) until 2005. This is a conservative measure of productivity that excludes book chapters and articles published in less internationally recognized journals; it is likely more accurate than self-reported productivity used by studies that rely on survey data, such as the Survey of Earned Doctorates. Using article counts as a measure of productivity is standard in the literature on scientific productivity; indeed, the six most recent papers on scientific productivity in the NBER working paper series rely on article counts. Article counts remain standard even as collaboration increases, because it is highly correlated with co-author-weighted publication counts (Cole and Zuckerman 1985; Wagner-Dobler 1997). We do, however, assess the robustness of our results to this alternative measure. We dismiss journal impact factor (JIF) weighted productivity measures because they confound quantity with quality – the two distinct outcomes that we expect IDR to affect differentially. We examine the number of articles at the person and person-year level in our analyses.

*Visibility.* We measure visibility by collecting the (forward) citations that have accrued to each published article (indexed in WoS) as of 2010. Citations to an individual paper are a more precise measure of a paper's impact than (the prestige of) its journal of publication; however, we control for journal impact factor in all models. In addition to this paper-level measure, we aggregate (by taking the mean) to obtain person and person-year level measures of visibility for certain analyses. When aggregated, both productivity and visibility are highly right-skewed, so we take the natural log as previous researchers have done (Allison and Long 1987; McBrier 2003; Prpic 2002) or use a count (negative binomial) model for models of productivity. To test Hypothesis 3, where we hypothesize about variability in citations, we calculate the standard deviation of a scientists' citations.

*Independent Variables*

*Interdisciplinary Research (IDR).* Our measure of IDR is borrowed from Porter and colleagues (2007:134). We measure interdisciplinary research at the paper level, aggregating up to the person or person-year level by taking the mean IDR score across each scientist's set of papers. Compared to other measures of diversity, Porter's measure of "integration" incorporates not only the variety (i.e., number) of categories and their balance (i.e., the evenness of the distribution), but also their similarity (i.e., their cognitive distance) into one index (Rafols and Meyer 2010). The categories of interest are the 244 WoS Subject Categories, or SCs (examples include Sociology, Chemical Engineering, and Organic Chemistry); the Web of Science assigns 1-6 SCs to each indexed journal, which we then extend to each constituent reference. As input for the measure of interdisciplinary research, we pool all SCs from the focal paper's set of references (rather than SCs of the focal paper itself, or SCs of papers that cite the focal paper), an approach that best gauges knowledge integration (Porter, Cohen, Roessner, and Perreault 2007:127). This captures the breadth of research referenced (and presumably integrated) in a paper. While the variety of SCs and their balance is specific to each focal paper's bibliography, their similarity ($s_{ij}$, from a SCxSC co-citation matrix which we convert to cosines)[3] is derived from the population of all WoS-indexed articles and thus shared by all focal papers. Porter's measure is a particular parameterization of the Sterling Index:

$$1- \Sigma_{ij} \ s_{ij} \ p_i \ p_j$$

where $s_{ij}$ is the similarity between SCs i and j, $p_i$ is the proportion of referenced papers in subject category i, and $p_j$ is the proportion of referenced papers in subject category j (Rafols and Meyer 2010: 267-8). We demonstrate the IDR calculation, step-by-step, for three hypothetical articles in Appendix Table A1. Intuitively, a paper's IDR score increases as it references more, relatively unrelated SCs[4] (Porter, Cohen, Roessner, and Perreault 2007:277). The IDR score ranges from 0 to 1, with scores closer to 1 indicating greater interdisciplinarity. For paper and person-year analyses, the IDR score is zero for 551 papers and 99 person-years. These values occur when the person has a very low productivity for that year (or has a single paper that references only one subject category). Given that these low productivity years obscure the production penalty that we hypothesize for IDR papers (because low IDR scores are associated with low productivity in the years where someone publishes little), we restrict all analyses to exclude observations with zero IDR scores. The visibility analyses are robust to including these observations; the productivity effects are curvilinear with this inclusion because IDR scores of zero only occur in low productivity years, and high IDR scores are also associated with low productivity (as we hypothesize). For the person-year analysis, we use the mean IDR score of papers published in that year to predict productivity.

~ Table A1 about here ~

---

[3] Off-diagonal elements in a SCxSC matrix for the year 2007 would indicate the number of papers that cited both SCs (i and j) between 2002 and 2006. To construct our SCxSC matrix, we summed three square matrices for the years 1987, 1997, and 2007 for the person-level analyses. We used the matrix closest to the year of publication for the person-year level analyses.

[4] The IDR score is only mildly (r=.15) related to the number of papers referenced. When we control for number of references in the paper and person-year analyses, the reported results are unchanged.

While a number of other measures of interdisciplinary research have been proposed and used in previous research, no other measure incorporates the relatedness of the categories that are joined. This is true for institution-level measures, e.g., the number of disciplines that members represent (Birnbaum 1981); person-level measures, e.g., whether an individual faculty member has a joint appointment (Jacobs and Frickel 2009); and even other publication-level measures. For example, Clemens and colleagues (1995:454) assess whether the paper is cited in a discipline other than the discipline the author(s) represented. Larivière and Gingras (2010) assess the percentage of a paper's cited references made to journals from other fields without considering the relatedness of those fields (e.g., whether a sociology paper references a cognate field like anthropology, or a more distant field like geology). The same limitation characterizes other research on knowledge products that span categories, including Hsu and colleagues' (2009) research on films and Fleming and colleagues' (2007) research on patents. In contrast, we rely on Porter's measure of IDR (which incorporates distance, or dissimilarity, between every pair of fields) and – to assess the extent to which distance matters above and beyond sheer variety (H4) – we compare it to the effects of a less refined measure: the total number of unique subject categories (SCs) appearing in a scholar's pooled set of references.

We ascertain the validity of this measure in a number of ways. First, by showing how the IDR score increases as a scientist references not only more, but more unrelated, fields, Table A1 provides face validity. Second, the IDR score (which is based on the variety, evenness, and distance of referenced SCs) is also positively correlated (r=0.18) with the number of SCs that characterize the focal paper itself (recall that a given paper is assigned 1-6 SCs). This is to be expected if focal paper SCs adequately capture the content of a paper that is more thoroughly gleaned from an analysis of the SCs it references, and it demonstrates convergent validity. Third, as we should expect, the mean IDR score of articles published in the interdisciplinary journals *Science*, *Nature*, and *PNAS* (0.71, 0.71, and 0.70, respectively) are significantly higher than the average IDR score in our sample (0.62) – more than half a standard deviation above. Thus we are reassured that the measure we use adequately captures the concept of interest to us.

**Field IDR**. We use both continuous and categorical measures of field-level IDR. The continuous measure is calculated as the average IDR score for the nine broad fields represented in our sample of scientists, including physics, chemistry, biology, social science, and most types of engineering (chemical, civil, material, electrical). The categorical measures indicate whether the scholar is in the life sciences (the highest IDR field in our sample) and whether the scholar is in electrical engineering (one of the lowest IDR fields in our sample). Data from the Survey of Earned Doctorates (using self-reports on the interdisciplinary nature of one's dissertation) confirm these as high (life sciences) and low (electrical engineering) interdisciplinary fields. Our data further corroborate this claim: 22% of the life science journals represented in our sample are interdisciplinary (i.e., their subject category is "Multidisciplinary"), compared to only 7% of the electrical engineering journals. We test H5 in two ways, at two levels. At the person-level, we compare two subsets of scholars: those working in a high IDR field (life sciences) and those working in a low IDR field (electrical engineering). At the person-year level, we use the continuous field level IDR score (the mean IDR score of all sampled scientists in each field), and interact that with the individual IDR score.

*Control Variables*

The scientific papers we analyze are nested within persons, so, when comparing publications across persons, we control for characteristics at this level - including gender, professional age, and status. Gender and professional age have been shown to influence engagement with IDR (Dahlander and Frederiksen 2012; Mansilla 2006; van Rijnsoever and Hessels 2011), as well as productivity and visibility (Leahey 2006; Maliniaka, Powers, and Walter 2013). Gender of each scientist was derived from analysis of first names, as well as information (pictures and pronouns) used on the scientists' websites. Professional age was calculated by subtracting the year of PhD receipt (obtained from CVs and Proquest Dissertation Abstracts) from 2005, the year publication data were collected. To proxy status of current institution, we use the number of faculty who are members of the National Academy of Sciences (collected from The Center for Measuring University Performance). To measure the quality of an individual scholar, we include the ranking of one's PhD-granting institution, obtained from the Academic

Ranking of World Universities, (http://www.arwu.org) and reverse-coded so that higher values indicate higher quality. Although certainly a rough measure of individual quality, it helps to rule out concerns about "smarter" scholars being more (or less) likely to engage in IDR research.

The papers we analyze are also nested within journals and fields, so we also incorporate controls at these levels. Because interdisciplinary research has been linked with the prestige of the journal in which it appears (Rinia, van Leeuwen, and van Raan 2002), we control for mean impact factor of the journals each scholar has published in. When modeling productivity, we control for average turn-around-time for journals in the field, obtained from Bjork and Solomon (2013). When modeling visibility, we control for average citations per paper in the field, obtained from Thomson Reuters. Because the relationship between interdisciplinary research and citation counts depends on the field (Hamilton, Narin, and Olivastro 2005; Larivière and Gingras 2010; Zitt, Ramanana-Rahary, and Bassecoulard 2005), and because we want to rule out increased audience size as an alternative explanation for heightened visibility, we control for field size (the number of PhDs produced in a recent year, obtained from NSF's Survey of Earned Doctorates). In addition, we control for the potential reach of each paper (Fleming & Sorenson 2004:919), captured by the number of SCs that classify the focal paper itself, rather than its references. We presume that papers classified in multiple fields will be brought to the attention of a larger body of scholars, potentially boosting citations (thus, this is another control for audience size). Although our sample is center-based, we did not include center level controls because none reached statistical significance and the intra-class correlation coefficient (0.11) suggested minimal clustering by center. In paper-level and person-year analyses, we controlled for additional variables that might influence visibility and/or productivity. Teams produce more highly cited papers, on average, than sole authors (Montpetit, Blais, and Foucault 2008; Wuchty, Jones, and Uzzi 2007), and authors who have worked together previously may face fewer production penalties compared to newly-formed collaborative teams. Thus, we included the number of authors on the paper and a binary variable indicating whether this combination of authors had published together before (in yearly analyses, these were measured as the average number of authors and the proportion of papers with repeat collaborators, respectively). We also control for both lagged cumulative publication experience as well as lagged cumulative citations (logged to remedy skewness). Finally, we controlled for publication year as older publications have more time to accrue citations than more recent publications.

*Statistical Approach*

Interdisciplinarity is measured at the paper-level, but some scientists' research outcomes (such as productivity) cannot be measured at the paper level, so we present models at the paper, person-year and person level of analysis. This allows us to model outcomes at the most appropriate level(s), incorporate level-specific control variables, and assess the robustness of our results.

Our first analysis takes place at the person level (n=854 scientists). We rely on path analytic techniques, which are ideal for two reasons. First, we have two main outcomes of interest (productivity and visibility), and path analysis allows us to model these outcomes and their own inter-relationship simultaneously, in the same model. Second, path models allow us to examine direct as well as indirect effects – of, for example, IDR→productivity→visibility – and thus explicitly model the influence of intervening variables, unlike a regression-based mediation analysis. In addition, the structural equation modeling (sem) package in Stata 13, which we use to estimate path models, is ideal for handling missing data. Rather than deleting data in a listwise fashion, which is the default strategy in most statistical packages, this package relies on a full information maximum likelihood estimation procedure. This strategy permits the inclusion of all available data (Anderson 1957) and bypasses the need to impute data.

We also take advantage of our data structure – panel data over the career of individual scientists – to estimate models at the paper and person-year levels of analysis. This allows us to assess the robustness of the person-level results, to better assess causal direction, explicitly test for field-level moderators, and examine the effects of paper-level variables. A Hausman test suggests a fixed-effects model is more appropriate than a random effects model ($\chi^2$= 405.57; $\chi^2$= 382.90; *p*<.0001 for productivity and visibility, respectively). We estimate fixed-effects regression models at the person-year level (n=8,779) -- because a person's productivity cannot be captured at the paper level – but also at the paper level (n=29,782) when modeling visibility. The fixed-effects models allow us to look at within-person variation on our dependent variables and control for unmeasured person-level characteristics, and thereby provide the most conservative test of our hypotheses. However, when testing Hypothesis 5 on field-level IDR, we estimate random-effect models with robust standard errors (for visibility) and population-averaged negative binomial models with robust standard errors (for productivity) because field IDR, and other field-level characteristics like journal turn-around-time and average field citations, are time-invariant and would be dropped from a fixed-effects model[5]. We use multiple imputation because of missing data for some variables; however, results are the same without multiple imputation.

RESULTS

We begin by describing our data and measures in Table 1. Panel A describes our sample of 854 scientists, all of whom have a publication record, making all our results reported herein conditional upon publication. Panel B provides descriptive statistics for all variables used in the person-year level models. There is ample variation on the key outcomes (visibility and productivity), both of which are left-skewed so we take the natural log for subsequent analyses. The key predictor, IDR, also displays variation -- ranging from 0.08 to 0.88 (on a scale of 0 to 1). To detect problems of multicollinearity, we calculated the variance inflation factors (VIF). In all models the VIF scores were below six, below the recommended cutoff value of ten (Neter, Wasserman, and Kutner, 1985). In the section on robustness checks below, we demonstrate the representativeness of our sample by comparing our sample of papers with the population of WoS papers analyzed by Uzzi et al. (2013) and comparing our sample of scientists with the population of PhD-level scientists from NSF's surveys.

~ Table 1 about here ~

---

[5] Productivity results are substantively the same when estimated using a random-effects negative binomial model.

As expected (H1), IDR depresses scholarly productivity (see Table 2, Model 1); this effect is statistically significant at the 5% level. Recalling that the outcome variable is logged,[6] the coefficient of -0.77 suggests that a 10% increase in IDR reduces productivity by 7.7% over one's career, controlling for professional age and other factors. This effect suggests that interdisciplinary scientists do indeed experience lower productivity. This productivity penalty holds, and even gets stronger (b= -0.99**, SE=0.39), even when we weight the article count by number of coauthors (such that a paper with two authors only contributes 0.5 to a scholar's productivity). We also examine IDR's effect on productivity at the person-year level (in which IDR scores are averaged, and publications are summed). Using a model with person fixed-effects as well as time-period controls[7] (Table 3, Model 1) allows us to rule out differences in individual propensities to engage in IDR. Here, too, IDR has a significant and negative effect (b= -0.143**) on productivity.  In those years when scholars do more interdisciplinary work, they publish fewer articles. Thus we find consistent support for a production penalty across both specifications, consistent with Hypothesis 1.

~ Tables 2 and 3 about here ~

However, once published IDR shines: in support of H2, we find that IDR increases scholarly visibility. The coefficient for IDR (+0.69*** in Table 2, Model 1) suggests that a 10% increase in IDR increases a scholar's citation, on average, by 6.9%. In Table 3, Model 4 (a paper level model with person and year fixed-effects), we see that IDR has a positive and significant effect on visibility. This effect also holds when we exclude 'group authors,' whose papers tend to be highly cited (results not shown). Because articles in multidisciplinary, high-impact journals like Science, Nature, and PNAS could be driving the visibility benefit, we omitted these 171 papers (0.05% of total) from all measures. All hypothesized results are robust to this change.[8]

---

[6] When the dependent variable has been log-transformed and the predictors have not, a one-unit increase in the independent variable produces a 100*(coefficient) percent change in the dependent variable.

[7] We cannot use year as well as person fixed effects when modeling productivity at the person-year level because variation on the dependent variable (i.e., # articles published in a year) is minimal.  Instead, we use 5-year time windows to control for time-trends in publication.

Although IDR is more visible, on average, we also find support for H3: having a record of interdisciplinary scholarship increases the overall variance of a scientist's paper (H3). In Table 2, Model 2, we see that scientists who publish more IDR are more likely to produce both frequently-cited <u>and</u> rarely-cited works. Here, we model the *standard deviation* in citations rather than mean citations (and also control for the standard deviation rather than the mean of journal impact factor). We find that interdisciplinary scientists, in addition to having a greater total number of citations, also experience more *variability* in citations across their papers. Scientists with a record of interdisciplinary scholarship experience more 'hits' and more 'flops' than their mono-disciplinary counterparts. To investigate this further, we examine whether it is indeed the high IDR papers that display more variability in citations (rather than highly interdisciplinary scientists having a few disciplinary papers that are not well cited). Using the median IDR score, we distinguish low IDR and high IDR papers. Then, for each scientist who had at least two papers of each type (n=647), we calculate the standard deviation of the citations received by their low IDR papers, and do the same for their high IDR papers. As theorized, the variance of scholars' high IDR papers (mean=30.4) is higher than the variance of their low IDR papers (mean=24.2), and this difference is statistically significant ($p<.0015$).

The distance (or cognitive dissimilarity) between fields contributes to the visibility benefit and determines the productivity penalty, providing support for H4. We assess this by extracting distance out of the IDR measure: we simply calculate the total number of unique SCs referenced by a scientist (across all his papers) with no regard for their similarity (range 1-107, mean=32). When we substitute this measure for the IDR measure (see Table 2, Model 3), the positive effect on citations holds, suggesting that even spanning related fields improves citations (if only slightly), presumably by broadening one's prospective audience. However, the negative effect on productivity does not hold. In fact, this alternative measure of IDR *positively* affects productivity, perhaps because drawing on multiple disciplines expands the number of possible journal outlets. Simply drawing upon more SCs doesn't hinder productivity, unless those SCs are cognitively dissimilar. This suggests that it is more difficult to produce and successfully publish scholarship that spans unrelated fields (e.g., chemical engineering and anthropology) than related fields (e.g., chemical engineering and civil engineering). In supplementary analyses (not reported), we confirm these results at the paper- and person- levels of analysis by controlling for number of SCs referenced in each paper and we confirm that our IDR measure (which incorporates distance) is the only significant predictor. This offers support for the mechanism behind the production penalty we theorize: IDR is cognitively difficult and slow to produce when it blends disparate fields.

---

[8] These journals are all classified into the "Multidisciplinary" subject category (SC). Another way to identify interdisciplinary journals is to examine those with multiple SCs assigned (e.g., the maximum is 6). At the journal level, we find that the number of SCs is negatively related to Journal Impact Factor. (-0.15), reassuring us that a correlation between interdisciplinary journals and impact is not driving the visibility benefit we document.

Lastly, we examine how the effects of IDR depend on the nature of the field. As expected (H5b), we find that in highly interdisciplinary fields like the life sciences, IDR's negative impact on productivity is weaker than in less interdisciplinary fields like electrical engineering; where in fact, IDR fails to reach statistical significance (Table 2, Models 4 and 5). These sub-analyses fail to support H5a (field differences in the reception benefit), perhaps due to small sub-sample sizes and low statistical power. As an alternative way of assessing the interaction between IDR and field IDR, we retain the full set of observations, and interact the continuous field-level IDR measure with the individual IDR measure. Results from models at the person year level (see Table 3) reveal that both field-level IDR and the interaction term have positive and significant effects on productivity (Model 3) and visibility (Model 6).[9] In other words, field-level IDR bolsters IDR's positive effect on citations, and lessens its negative effect on productivity (in fact, the relatively flat line in Figure 1 suggests that scholars in high IDR fields do not face a productivity penalty). High IDR fields are more receptive to IDR work, and thereby invoke fewer penalties for producing this type of work (perhaps they provide better training in how to manage the cognitive challenges, and/or are more amenable to IDR in the peer review process). These effects are depicted in Figures 1 and 2, in which results from Table 3 (Models 3 and 6) are used to plot standardized individual IDR scores for scholars in an average IDR field and those in fields that are one standard deviation above and below the mean. Overall, results from the field-specific sub-analyses (comparing life sciences and electrical engineering) and the interaction of IDR and field-level IDR provide support for Hypothesis H5b and mixed support for Hypothesis H5a.

---

[9] Given our interest in field-level effects, and because field-level IDR is time-invariant, fixed effects models cannot be estimated. Instead, we use robust standard errors and a set of controls to account for individual-level differences in propensity to conduct IDR research.

~ Figures 1 and 2 about here ~

All of the effects noted above hold even in the face of important controls. In the person-level path analytic models, we controlled for precursors of IDR that have been identified by others, including gender, professional age, and status. Perhaps most surprising is the effect of gender on IDR: contrary to widely held perceptions and some previous empirical research (Rhoten and Pfirman 2007), we find that women are not more likely than men to engage in IDR. We also find support for Phillips and Zuckerman's (2001) middle status conformity finding. Using the number of National Academy of Science (NAS) members at one's institution as a measure of status, we find the expected inverted U-shape relationship between status and IDR: as Model 1 (Table 2) shows, the main effect of status is negative, and the squared term is positive. This suggests that scientists at both low and high status universities engage in IDR; scientists at middle-status universities are in a precarious position (where "the prospect of classification as a full-fledged player and the threat of delegitimation both loom large" (Phillips and Zuckerman 2001: 384)) and thus opt to conform to a disciplinary tradition. At the journal and field level, we controlled for variables that likely impact productivity and/or visibility – including field size, average citations per paper in the field, average turn-around time at journals in the field – as well as lagged productivity (when modeling productivity as well as visibility) and lagged visibility (when modeling visibility as well as productivity). While they all have intuitive and often significant effects, they do not alter (or render insignificant) the main findings.

*Drivers of the Productivity Penalty*

    <u>Cognitive and Collaborative Challenges.</u> We theorize that IDR projects typically experience a steep learning curve as well as communication challenges with diverse collaborators. Supplemental data sources and analyses lead us to conclude that IDR projects indeed face these communicative and collaborate hurdles.

    First, we examine the interaction between repeat collaboration and IDR to see whether collaborators learn to work together better over time. Indeed, working repeatedly with a similar set of collaborators reduces the productivity penalty (b=1.51*, se=.077).

    Second, we capitalize on survey data that we collected from a subset of scholars in the archival data. Scholars were asked about the nature of the collaboration on their most recent co-authored paper, and we match these responses to the IDR scores for these papers (n=68). The IDR scores of papers by these scholars are not significantly different from the rest of our sample. Despite the small sub-sample size, a series of t-tests reveal marginally significant challenges associated with the production of IDR. Interdisciplinary collaborations are fraught with communication difficulties: communication is reportedly less clear (p=.065), more difficult (p=.083), and of lower quality (p=.109). Moreover, interdisciplinary teams have more difficulty generating ideas (p=.103).

    And last, we create an alternative measure that captures multidisciplinarity rather than interdisciplinarity. Our theory suggests that it is challenging to incorporate distant ideas within a single paper. An alternative is that a scientist writes papers across different disciplines but each paper is mono-disciplinary; this scientist is multi- but not inter- disciplinary. Although multidisciplinarity may require additional expertise and a diverse network of collaborators, it does not require integration of diverse fields or coordination of diverse collaborators. To capture multidisciplinarity, we pool all of the subject categories across a person's publication record and then calculate IDR. This differs from a scientist's IDR

score, which is first calculated at the paper level, and then averaged across his set of papers. Consistent with our theorizing, scientists conducting multi-disciplinary research are *more* productive (b=0.58* SE=0.31); it is only scientists publishing interdisciplinary research who are *less* productive.

Challenging Review Process. Another possibility is that IDR work is penalized by reviewers and editors in the review process, but we find little support for this explanation. In order to examine this possibility, we collected data on the length of review process (i.e., turn-around time) for two journals represented in our data: one publishing articles with an above-average IDR score (0.64), and another publishing articles with a below-average IDR score (0.59). For the 711 articles published in these journals (written by 145 of our sampled authors), the median time under review is 85 days. However, there is no significant correlation between time under review and the paper's IDR score, and the IDR score does not predict time under review.

Confidence in this finding is buttressed by an analysis of data on unpublished working papers that we collected from www.ArXiv.org. We searched for working papers by the 854 authors in our sample, and were able to locate 220 papers written by 63 authors (see Table A4). Stripping all references from these working papers, identifying each referenced journal, and matching the journals with WoS SCs allowed us to calculate IDR scores for each working paper. Comparisons of the 220 working papers with the 3983 published papers by the same authors reveal that working papers are, indeed, more interdisciplinary, and this may make them more likely to be file-drawered or rejected, thereby contributing to the productivity penalty. But because working papers are more recent (average posting date in ArXiv.org is 2009, compared to an average publication date of 1997 for the published papers), and may be more interdisciplinary simply given the upward trend in IDR (documented by Porter and Rafols 2012, and confirmed with our data in Figure 3), it is difficult to interpret this difference. As a more stringent test, we took a closer look at our sample of working papers. For each working paper, we identified which papers were subsequently published (using author updates on www.ArXiv.org and online searches for each paper). We compared working papers that eventually got published with those that did not. Here, the eventually published papers (n=122) are actually more interdisciplinary than the still unpublished papers (n=115), a difference significant at p=.054 (two-tailed), suggesting that IDR papers are not hindered in the review process.

~ Table A4 and Figure 3 about here ~

Taken together, these supplementary data and analyses support our theorizing about cognitive and collaborative challenges associated with IDR research. The second stage of production – the review stage – may not be such a roadblock; we do not find IDR work more likely to be file-drawered or rejected. Rather, the first stage of production, where authors plan, coordinate, and conduct their research is the largest hurdle for interdisciplinary work and the main source of the production penalty. Although we acknowledge that these are not causal tests but associations, we subscribe to the view that documenting the role of a mechanism (or two) empirically strengthens claims of causal connections (Gross 2009; Reskin 2003).

*Robustness Checks and Selection Issues*

<u>Sample selection concerns.</u>  Additional analyses alleviate concerns that center affiliation is driving the results we report. It is true that our sample of academic scientists (described in Table 1), while large and heterogeneous in terms of represented fields, includes only scientists affiliated with at least one research center: the IUCRCs we study. But we can leverage our longitudinal data, which includes papers written by these scientists before they joined a center. Even after restricting the sample to papers published 5 years before, and 5 years after center founding and estimating fixed-effects models (Table A3, Models 1 and 2), the main effects of IDR (positive on visibility; negative on productivity) remain significant. Although descriptively we find that scientists are slightly more interdisciplinary in their publications after center formation, this could be attributed to the trend toward IDR (Figure 3). The interaction between IDR and center affiliation is not significant, suggesting that IDR work does not become more visible after joining a center (Model 1) and IDR work does not become less difficult (Model 2). This analysis suggests that center affiliation does not increase the benefits or reduce the cost of conducting IDR.

~ Table A3 about here ~

Our center-based sample is distinctive in some ways, but not in ways that alter the results we report. Compared to their counterparts unaffiliated with centers, university research center scientists tend to be more experienced (Bozeman et al. 2001) and productive (Biancani et al. 2014), and this may be particularly true when scholars are connected to industry (Carayol and Thi 2005) via IUCRCs. Indeed, we find that compared to the broader population of scientists in related fields (computer science, math, life sciences, physical sciences, social sciences, and engineering) represented in NSF's surveys of PhD-level scientists[10], our sample of scientists is older, more predominantly male, and more productive (see Appendix Table A2). Concerns about the older age of our sample are alleviated by examining subsets of our data. When we restrict our sample to youngest 20% of our sample (not more than 13 years post-PhD) in the person-level analysis, and when we restrict our sample to the first five years of a scientist's career in the paper-level analysis, our results hold. An interaction between IDR and age is negative and significant (b= -0.012***) suggesting that, if anything, the penalty is stronger for older scientists (but the size of the coefficient is very small). Concerns about the preponderance of men in our sample are alleviated when interactions between female and IDR fail to reach statistical significance, suggesting that results hold for both men and women. Concerns about our especially productive sample of scientists are alleviated when we find evidence of a productive penalty in other data sources: interdisciplinary scholars (measured as those who have received their PhD in one field but work in another) report significantly fewer publications than scholars whose fields of training and work correspond in The National Survey of Postsecondary Faculty (2003). Moreover, Biancani et al. (2014) find that when Stanford faculty take a joint appointment (one measure of interdisciplinary engagement), their productivity declines.

---

[10] For the comparison reported here, we compare our sample of scientists to scientists represented in the SED/SDR SESTAT data in 1998-2003.

~ Table A2 about here ~

And in other ways, our sample is representative of scientists, and scientific papers more broadly. Indeed, center affiliation is relatively common: almost one-third (32%) of faculty at extensive (R1) research universities nationwide are affiliated with a research center (Boardman and Corley 2008). And even if center-based scholars are more interdisciplinary (Ponomariov and Boardman 2010) and, as documented above, more productive, this would compress the 'true' productivity penalty, and make our test conservative. Although we cannot calculate our measure of IDR for PhD-level scientists in NSF's survey data, we can compare our sample of papers to the population of papers in Web of Science analyzed by Uzzi et al. (2013). Uzzi generously shared his measures for our sample of papers, so we can make direct comparisons in terms of both conventionality (a paper's tendency to reference journals that are commonly referenced together) and novelty (a paper's tendency to reference unusual journal pairings). The second panel of Table A2 shows that in the 1980s and 1990s, our papers are almost identical to the population of Web of Science papers in terms of conventionality. And, importantly, Table A2 also shows that our sample of papers exhibits slightly less novelty than the population, allaying concerns that center-based scholars are innovative superstars.[11] This comparison demonstrates that they are not exceptional, and our results are likely to apply to academic science more broadly. That said, and despite our additional efforts, we acknowledge that with our research design, we cannot completely rule out that there is something distinctive about center-based scientists.

Quality of IDR papers. We also find evidence suggesting that interdisciplinary papers are not distinctive in terms of their inherent quality. If IDR papers are lower in quality compared to other papers, this could explain the productivity penalty we document. If IDR papers are higher quality papers, this could explain the visibility benefit we document. Additional analyses allow us to rule out these possibilities. We control for potential quality differences at the individual-, journal-, and field-level.

---

[11] Appendix Table A3, Model 3, also replicates the Conventionality+Novelty effect from Uzzi et al. (2013) for our sample and demonstrates that IDR has an independent and significant effect on visibility.

Recall that we specify models with person fixed effects that control for unmeasured attributes, so we can be certain that the effects we document are not attributable to individual differences in aptitude or in propensity to engage in IDR. In other words, low quality scholars are not attracted to IDR (which could explain the productivity penalty), nor are high quality scholars (which could explain the reception benefit). When we examine field-level moderators, person fixed-effects are not possible (as in Table 3, Models 3 and 6), so we control for the quality of scientists' PhD institutions and their current institution. With respect to the quality of the journal, we do see a positive correlation between IDR and journal impact (r=0.25 in Table 1). But the positive effect of IDR on citations holds even when we control for journal impact (Table 2, Model 1), when we examine journal fixed-effects (results not shown), and when we restrict our analysis to the subset of low-impact journals -- those whose impact factor falls in the lowest quartile. In other words, high IDR papers in low-impact journals are more highly cited than other papers in those journals. Moreover, we found no difference in turn-around times between papers published in a high IDR journal and in a low IDR journal (see above). Thus, we doubt that IDR papers are inherently higher quality (as then they would likely move more quickly through the review process); we also doubt that they are lower quality (as then they would likely spend more time in development in the review process or be more likely to be rejected). These robustness checks are consistent with our argument that IDR, and not some unobserved heterogeneity, increases visibility and reduces productivity.

DISCUSSION

The main contribution of this paper is to empirically investigate the potential costs, as well as the widely touted benefits, of interdisciplinary scholarship. To do this, we collected and collated data from various sources for a sample of over 850 center-based scientists and their 32,000 publications, to provide the first systematic and mid-scale assessment of IDR's impact. Our results demonstrate that indeed, IDR benefits scientists: it improves their visibility in the scientific community, as indicated by cumulative citation counts. But we also document a productivity penalty associated with IDR: it depresses the number of articles that scientists publish. We see these effects in analyses of papers (i.e., high IDR papers are more highly cited), in yearly analysis (e.g., scientists publishing high IDR papers publish less in that year), and at the person-level (e.g., scientists who publish IDR have higher citations and higher variation in their citations). And when compared, the productivity penalty for IDR (standardized coefficient for direct effect = -0.06) outweighs the reception benefit (standardized coefficient for direct effect = +0.04) of engaging in this research. In other words, engaging in interdisciplinary research depresses productivity more than it increases citations. Specifically, compared to a scholar in the 20[th] percentile of IDR, a scholar in the 80th percentile of IDR produces 2 fewer articles (24, rather than 26) and garners almost 50 more citations (297, rather than 253).[12] The productivity penalty is strong enough to make the *total* effect of IDR on citations – i.e., the direct effects reported above, plus the indirect effect of IDR on citations via productivity – slightly negative (standardized coefficient for total effect = -0.28). Apparently the learning curve is steep: it takes more time, effort, diligence, and perhaps coordination to master (at least aspects of) different fields and to work with scientists trained in disparate disciplines. That said, we cannot assess whether this trade-off is harmful or beneficial for a scientist (do 50 more citations offset 2 fewer publications?). We can say that the greater visibility that this work receives is accompanied by fewer publications published.

We also make several theoretical contributions to the literature. First, we re-orient away from reception penalties (which dominate the categories literature) toward production penalties, and explore a new form of production penalty. The few recent papers that examine production penalties focus on reduced quality (Kovács and Johnson in press), documenting through blind taste tests (i.e., controlling for audience perception), for example, that category-spanning wines are rated more negatively -- presumably because of skill deficiencies on the part of the winemakers (Negro and Leung 2013). In contrast, we focus on another form of production penalty that besets category-spanning work, at least in in the realm of science: reduced productivity. Supplementary data on unpublished working papers, journal turn-around times, and a survey of authors' experiences, in addition to supplementary models incorporating individual fixed effects, suggest that productivity is hampered by communication and coordination hurdles faced in the research process, and not by the review process. This needs to be reconciled with the implicit production benefits found in the innovation and diversity literatures: better decisions (Beckman and Haunschild 2002), heightened creativity (Pelled, Eisenhardt, and Xin 1999), and organizational centrality (Powell, Koput, and Smith-Doerr 1996). Our work suggests a potential short-run (fewer papers) and long-run (higher visibility) tradeoff.

---

[12] These predicted values are obtained from Table 2, Model 1. We used the 'predict' post-estimation command in Stata, then calculate the average prediction for each decile of IDR (which we then exponentiated because the outcomes in our analysis were logged).

Second, we push research forward by demonstrating that production penalties and reception benefits depend on characteristics of the field. The production penalty (fewer papers) is more severe in a low IDR field (like electrical engineering) than in a high IDR field (like life sciences). This suggests that the penalties of doing IDR work are not the same across fields: the cognitive and collaborative challenges can be reduced through training, or peer review processes that are amenable to IDR papers may help develop that work. More broadly, the effect of category-spanning depends on the typicality of category-spanning in the field (see also Lo and Kennedy, in press). For example, it may be that these high IDR fields are shaped by high status actors who reap the benefit of category-spanning and normalize it for those who follow (Rao, Monin, and Durand, 2005; Sgourev and Althuizen, 2014).

Third, we move beyond mere category-spanning to take into consideration the relationship between the spanned categories. Certainly, spanning two dissimilar entities is qualitatively different from spanning two similar entities, but extant research on both category-spanning (Hsu, Hannan, and Kocak 2009; Negro, Hannan, and Rao 2010) and recombinant innovation (Fleming, Mingo, and Chen 2007) largely ignores this [but see Leahey & Moody 2014, and Kovacs and Johnson 2014]. In contrast, the measure of IDR we use is sensitive to such differences, allowing connections between two unrelated disciplines to contribute more to the IDR score than connections between two related disciplines. The distinction is crucial and allows us to distinguish between the effects of variety (i.e., branching out into a number of other fields) and distance (i.e., branching out into cognitively distant and unrelated fields). The distance measure we use is consistent with the logic of diversity and brokerage – where networks with structural holes are presumed to contain more distant, non-redundant knowledge – but here we actually measure distance in knowledge space. When we measure IDR as sheer variety (neglecting distance between fields), we find that the visibility benefit holds (papers that span dissimilar and even similar subfields receive more citations), but the productivity penalty does not (suggesting that the cognitive challenges and coordination costs associated with producing IDR are largely a function of the cognitive distance among fields). This is also the case when we calculate a measure of multidisciplinarity (which accounts for distance between fields represented in a scholar's oeuvre, but not between fields represented in a given paper). Taken together, these results suggest that efforts to integrate, in a single paper, the cognitive distance across disparate fields contributes to the reception benefit (i.e., greater citations) and drives the productivity penalty (i.e., fewer publications).

Investigating academic science also allows us to extend Pontikes' (2012) insight about the nature of the audience to reconcile an even broader array of literatures than she recognizes. Like Pontikes suggests (2012:111), scientists are likely 'market-makers,' eager to identify and develop innovative, game-changing ideas, who are thus drawn to (rather than repelled by) multi-category offerings like IDR. And this, of course, is what we find: interdisciplinary scientists do not experience the reception side penalties (e.g., devalued, overlooked) that the category-spanning literature has widely documented among 'market-takers.' Rather, as the innovation literature suggests (without explicitly referring to it as such), scientists derive a reception-side benefit: greater visibility. However, we also confirm that IDR is a high-risk, high-reward proposition, as indicated by both more overall citations and higher variation of citations; we confirm that innovative ideas have longer tails (Singh and Fleming, 2007). Thus, Pontikes' focus on the nature of the audience places an important scope condition on not just the category-spanning literature, but also the innovation and brokerage literature – which has implicitly focused on market-makers and reception-side benefits. Given that the nature of the audience matters, it would behoove scholars of both category-spanning and innovation to identify more explicitly whether the actors under study are market-takers or market-makers, and to consider how this shapes both reception benefits (such as visibility) and risks (such as variance).

Because this is the first mid-scale and empirical assessment of IDR's effects on scientists' careers, extensions will be fruitful. Most illuminating might be a closer examination of the audience. The audience does not just cite and thereby contribute to a cumulative citation count. Audience members themselves come from disciplinary (or interdisciplinary) homes. In this paper we focused on the distribution of (and distance between) *cited* disciplines: those represented in scholars' bibliographies. But a more qualitative examination of the cited disciplines themselves, and a comparison of such referenced disciplines with the *citing* disciplines (that is, the disciplinary homes of scientists who reference a given paper), would allow us to assess whether certain fields (or combinations of fields) have broader appeal than others. In concurrent analyses, we find that papers with high IDR scores are likely to be cited by papers that themselves are interdisciplinary. This preliminary analysis suggests that breadth breeds further breadth, but further exploration is warranted.

We also encourage extensions to broader samples and the utilization of alternative research designs. Although we studied 850 scientists from a wide range of fields and university settings, they are all affiliated with university research centers that foster connections with industry. Our analyses of potential selection biases and endogeneity concerns suggest that center affiliation does not drive the results we report here. However, a prospective research design that follows a large sample or population of academic scientists through their careers (as they move in and out of center affiliations and other interdisciplinary ventures) would complement our search for mechanisms and evidence consistent with our theorizing. Efforts underway to link NSF's Survey of Doctorate Recipients (SDR) to both Web of Science records and the NBER Patent database would be ideal in this regard.

Practically, should university research administrators and federal agencies like the National Academies of Science and the National Science Foundation continue to invest in IDR? Given that IDR scholars produce useful and noteworthy research that has more impact on the scientific community, the enthusiasm for IDR is not premature. However, our analysis suggests that it is not attuned to the

implications – especially the negative implications – that IDR has for individuals. Scholars who produce IDR work may be more likely to publish in top tier journals (the correlation between IDR and journal impact factor is 0.25), but their overall productivity is hampered. There is a clear production penalty associated with IDR work. Even though our analysis of working papers suggests that IDR does not hamper (and indeed seems to help) subsequent publication, additional analysis on a larger sample is warranted. The penalties of IDR may be more far-reaching than we document here.

To this point, we encourage more direct examination of IDR's impact on scientific careers. We do not examine likelihood of receiving tenure, so we cannot assess whether the reduced productivity that IDR induces is offset by increased citations. This would be an important and useful extension, especially as universities re-organize to train scientists to be interdisciplinary. It is important to understand whether IDR enhances or damages career prospects for those starting out their careers. Certainly it appears that interdisciplinary scholars bear additional burdens: they struggle to master multiple domains of knowledge, to integrate them in a single work, and to coordinate with coauthors from different backgrounds. But these projects also appear to be received well by fellow scientists. More focused examination of the career outcomes of individual scientists is still needed. In addition, future research should examine forms of productivity other than publishing (e.g., patenting, consulting, advising) in order to examine whether scientists are compensating for their fewer publications with other activities.

Given the continued interest and enthusiasm for interdisciplinarity (National Research Council 2014), it is important that the costs and benefits of this type of research be evaluated empirically. We have attempted a step in that direction. Even if the individual level costs of such work are substantial, the societal level benefits – in the form of more useful and valuable science – seem clear. This suggests that scientists and the scientific community need to reassess how to evaluate scholarship if the scientists are to be encouraged to continue to engage in interdisciplinary research.

**Figure 1. Impact of IDR on Scientists' Yearly Productivity by Field**

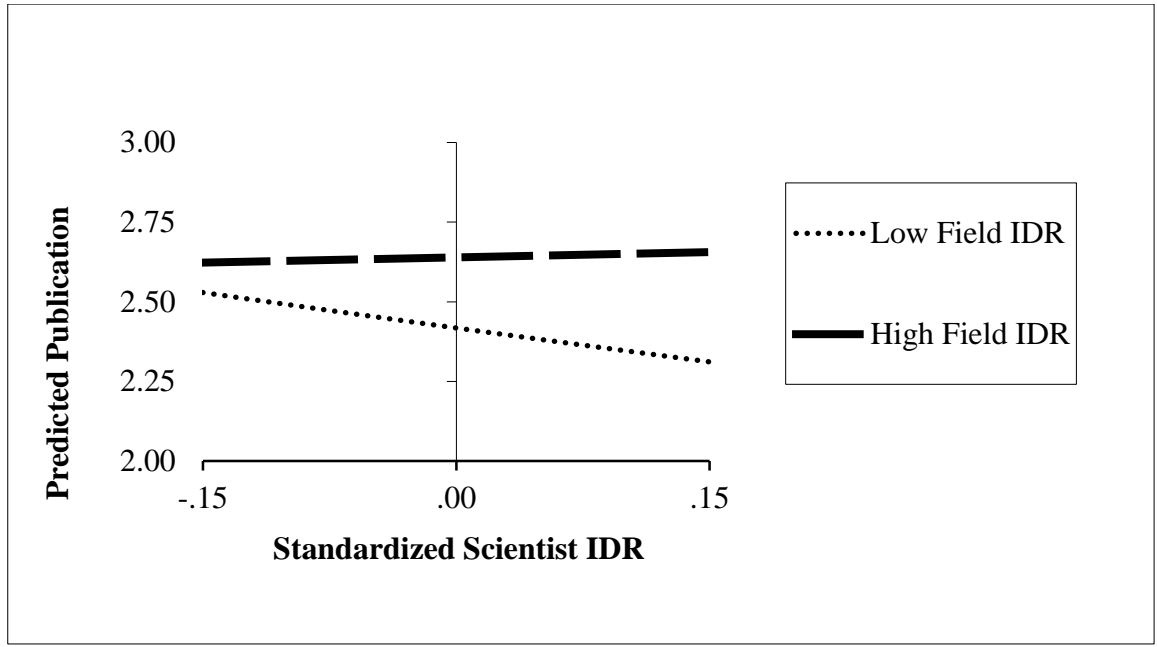

**Figure 2. Impact of IDR on Paper Visibility by Field**

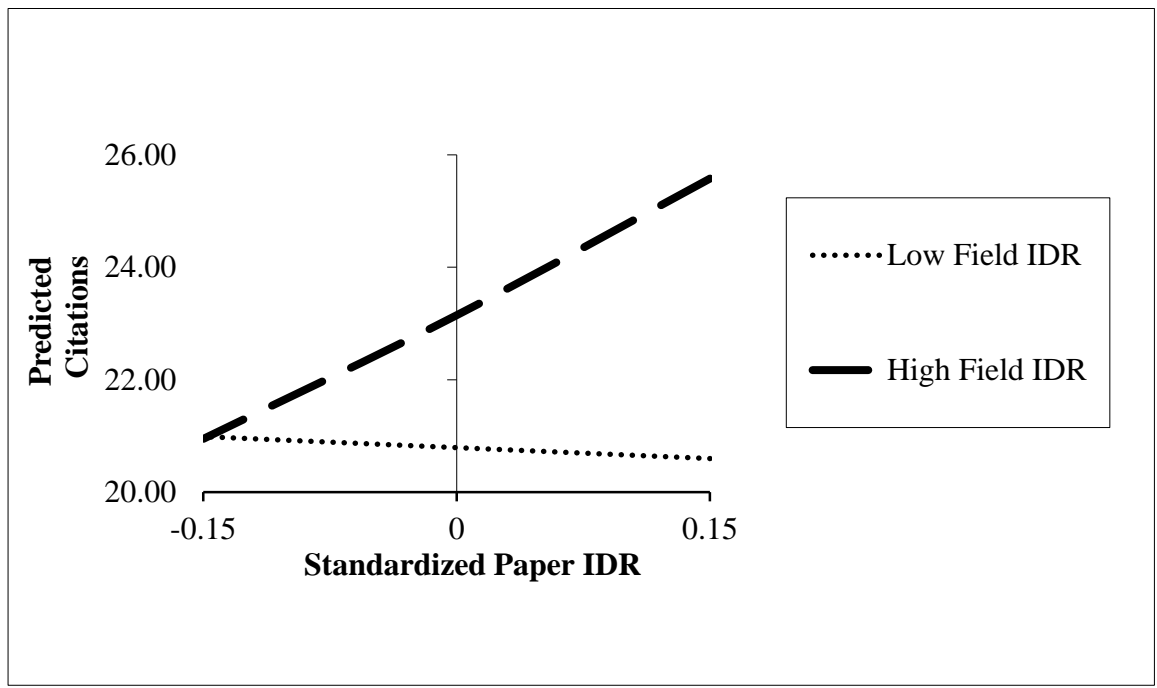

**Figure 3. Trend in IDR over time**

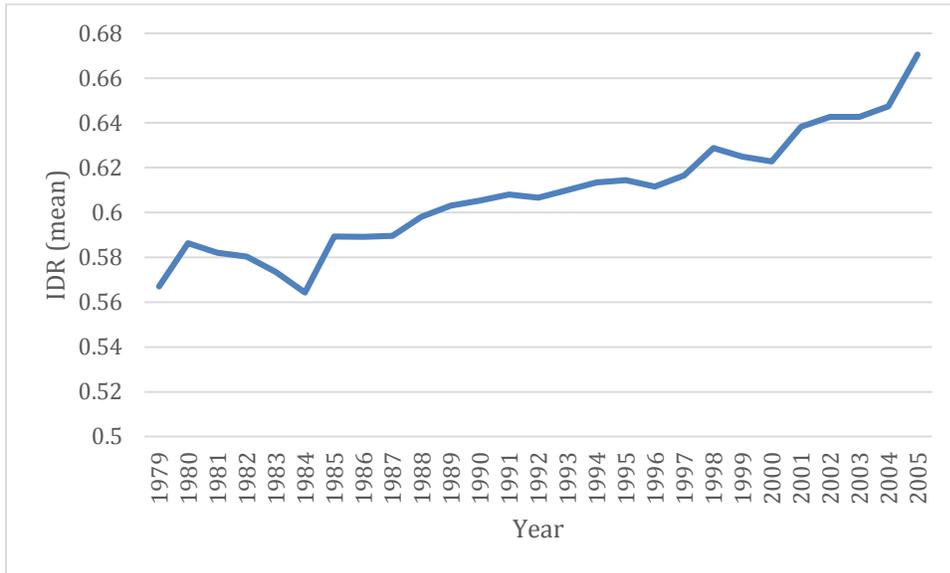

Note: The data used to produce this figure are described in detail in the text. See page 10-11 for a description of the sample of scholars and their ISI Web of Science publications, and page 12-13 for a description of the IDR measure (Porter et al. 2007).

# REFERENCES

Abbott, Andrew. 2001. *Chaos of Disciplines*. Chicago: University of Chicago Press.

Allison, Paul D. and J. Scott Long. 1987. "Interuniversity Mobility of Academic Scientists." *American Sociological Review* 52:643-652.

Anderson, T. W. 1957. "Maximum Likelihood Estimates for Multivariate Normal Distribution when Some Observations are Missing." *Journal of the American Statistical Association* 52:200-203.

Beckman, Christine M and Pamela R Haunschild. 2002. "Network learning: The effects of partners' heterogeneity of experience on corporate acquisitions." *Administrative Science Quarterly* 47:92-124.

Beckman, Christine M. 2006. "The Influence of Founding Team Company Affiliations on Firm Behavior." *Academy of Management Journal* 49:741-758.

Biancani, Susan, Daniel A. McFarland, and Linus Dahlander. in press. "The Semiformal Organization." vol. Published online in Articles in Advance 04 Feb 2014. Organization Science.

Birnbaum, Philip H. 1981. "Integration and Specialization in Academic Research." *The Academy of Management Journal* 24:487-503.

Boardman, Craig and Barry Bozeman. 2007. "Role Strain in University Research Centers " *Journal of Higher Education* 78:430-463.

Braun, Tibor and Andreas Schubert. 2003. "A Quantitative View on the Coming of Age of Interdisciplinarity in the Sciences: 1980-1999." *Scientometrics* 58:183-189.

Brint, Steven. 2005. "Creating the Future: 'New Directions' in American Research Universities." *Minerva* 43:23-50.

Burt, Ronald S. 2004. "Structural Holes and Good Ideas." *American Journal of Sociology* 110:349-399.

Carayol, Nicolas and Thuc Uyen Nguyen Thi. 2005. "Why do academic scientists engage in interdisciplinary research?" *Research evaluation* 14:70-79.

Carnabuci, Gianluca and Jeroen Bruggerman. 2009. "Knowledge Specialization, Knowledge Brokerage, and the Uneven Growth of Knowledge Domains." *Social Forces* 88:607-642.

Carroll, Glenn R. 1985. "Concentration and Specialization: Dynamics of Niche Width in Populations of Organizations." *American Journal of Sociology* 90:1262-1283.

Clemens, Elisabeth S., Walter W. Powell, Kris McIlwaine, and Dina Okamoto. 1995. "Careers In Print: Books, Journals, and Scholarly Reputation." *American Journal of Sociology* 101:433-494.

Cummings, Jonathon N. and Sara Kiesler. 2005. "Collaborative Research across Disciplinary and Organizational Boundaries." *Social Studies of Science* 35:703-722.

—. 2007. "Coordination costs and project outcomes in multi-university collaborations." *Research Policy* 36:1620-1634.

Dahlander, Linus and Lars Frederiksen. 2012. "The Core and Cosmopolitans: A Relational View of Innovation in User Communities." *Organization Science* 23:988-1007.

Diamond, Arthur M. 1986. "What is A Citation Worth?" *Journal of Human Resources* 21:200-215.

Evans, James A. 2008. "Electronic Publication and the Narrowing of Science and Scholarship." *Science* 321:395-399.

Ferber, Marianne. 1986. "Citations: Are They an Objective Measure of Scholarly Merit?" *Signs* 11:381-389.

Fleming, Lee. 2001. "Recombinant Uncertainty in Technological Search." *Management Science* 47:117-132.

Fleming, Lee, Santiago Mingo, and David Chen. 2007. "Collaborative Brokerage, Generative Creativity, and Creative Success." *Administrative Science Quarterly* 52:443-475.

Fleming, Lee and David M. Waguespack. 2007. "Brokerage, Boundary Spanning, and Leadership in Open Innovation Communities." *Organization Science* 18:165-180.

Freeman, John and Michael T. Hannan. 1983. "Niche width and the dynamics of organizational populations." *American Journal of Sociology* 88:1116-1145.

Gans, Joshua S. and Fiona Murray. 2014. "Credit History: The Changing Nature of Scientific Credit" in A. Jaffe and B. Jones, eds., The Changing " in *The Changing Frontier: Rethinking Science and Innovation Policy*, edited by A. Jaffe and B. Jones. Chicago, IL: University of Chicago Press.

Hamilton, KS, F Narin, and D Olivastro. 2005. "Using bibliometrics to measure multidisciplinarity." Westmont, NJ: ipIQ, Inc.

Hannan, Michael T and John H Freeman. 1989. *Organizational Ecology*. Cambridge, MA: Harvard University Press.

Hannan, Michael T. and Balazs Kovacs. 2011. "Category Spanning, Distance, and Appeal." working paper: Stanford GSB Research Paper 2081.

Hannan, Michael T., Laszlo Polos, and Glenn R. Carroll. 2007. *Logics of Organization Theory: Audiences, Codes, and Ecologies*. Princeton, NJ: Princeton University Press.

Hargadon, Andrew B. 2002. "Brokering Knowledge: Linking Learning and Innovation." *Research in Organizational Behavior* 24:41-85.

Harrison, David A. and Katherine J. Klein. 2007. "What's the Difference? Diversity Constructs as Separation, Variety, or Disparity in Organizations." *Academy of Management Review* 32:1199-1228.

Hsu, Greta. 2006a. "Evaluative Schemas and the Attention of Critics in the US Film Industry." *Industrial and Coprorate Change* 15:467-496.

—. 2006b. "Jack of All Trades and Master of None: Audences' Reactions to Spanning Genres in Feature FIlm Production." *Administrative Science Quarterly* 51:420-450.

Hsu, Greta, Michael T. Hannan, and Ozgecan Kocak. 2009. "Multiple Category Memberships in Markets: An Integrative Theory and Two Empirical Tests." *American Sociological Review* 74:150-169.

Jacobs, Jerry A. and Scott Frickel. 2009. "Interdisciplinarity: A Critical Assessment." *Annual Review of Sociology* 35:43-65.

Knorr Cetina, Karin. 1980. "The Scientist as an Analogical Reasoner." in *The Social Process of Scientific Investigation*, edited by K. Knorr-Cetina, R. G. Krohn, and R. Whitley. Boston: Kluwer.

Kovacs, Balasz and Michael T. Hannan. 2011. "Category spanning, distance, and appeal." Graduate School of Business, Stanford University.

Kovacs, Balazs and Amanda Sharkey. 2014. "Paradoxical Publicity: How Status Can Lead to Lower Perceived Quality." *Administrative Science Quarterly* 59:1-33.

Kovács, Balázs and Rebeka Johnson. in press. "Contrasting alternative explanations for the negative effects of category spanning: A study of restaurant reviews and menus in San Francisco." *Strategic Organization*.

Lamont, Michele. 2009. *How Professors Think: Inside the Curious World of Academic Judgment*. Cambridge, MA: Harvard University Press.

Lamont, Michele, Gregoire Mallard, and Joshua Guetzkow. 2006. "Beyond Blind Faith: Overcoming Obstacles to Interdisciplinary Evaluation." *Research Evaluation* 15:43-55.

Larivière, Vincent and Yves Gingras. 2010. "On the relationship between interdisciplinarity and scientific impact." *Journal of the American Society for Information Science and Technology* 61:126-131.

Latour, Bruno. 1987. *Science in Action: How to Follow Scientists and Engineers through Society*. Cambridge, MA: Harvard University Press.

Leahey, Erin. 2006. "Gender Differences in Productivity: Research Specialization as a Missing Link." *Gender & Society* 20:754-780.

—. 2007. "Not by Productivity Alone: How Visibility and Specialization Contribute to Academic Earnings." *American Sociological Review* 72:533-561.

Leahey, Erin and James Moody. 2014. "Sociological Innovation through Subfield Integration." *Social Currents* 1:228-256.

Lingo, E.L. and S. O'Mahony. 2010. "Nexus work: Brokerage on creative projects." *Administrative Science Quarterly* 55:47.

Lo, Jade Yu-Chieh and Mark T. Kennedy. in press. "Approval in Nanotechnology Patents: How Institutional Logics and Pattern Repetition Affect Reactions to Category-blending." *Organization Science*.

Lynn, Freda B. 2014. "Diffusing through Disciplines: Insiders, Outsiders, and Socially Influenced Citation Behavior." *Social Forces* 93:355-382.

Maliniaka, Daniel, Ryan Powers, and Barbara F. Walter. 2013. "The Gender Citation Gap in International Relations." *International Organization* 67:889-922.

Mansilla, Veronica Boix. 2006. "Assessing Expert Interdisciplinary Work at the Frontier: An Empirical Exploration." *Research Evaluation* 15:17-29.

McBrier, Debra Branch. 2003. "Gender and Career Dynamics within a Segmented Professional Labor Market: The Case of Law Academia." *Social Forces* 81:1201-1266.

McPherson, Miller, Lynn Smith-Lovin, and James M. Cook. 2001. "Birds of a Feather: Homophily in Social Networks." *Annual Review of Sociology* 27:415-444.

Merton, Robert K. . 1977. "The Sociology of Science: An Episodic Memoir." Pp. 3–144 in *The Sociology of Science in Europe*, edited by R. K. Merton and J. Gaston. Carbondale, IL: Southern Illinois Press.

Millar, Morgan M. and Don A. Dillman. 2012. "Trends in Interdisciplinary Dissertation Research: An Analysis of the Survey of Earned Doctorates." NCSES Working paper #12-200.

Montpetit, Éric, André Blais, and Martial Foucault. 2008. "What Does it Take for a Canadian Political Scientist to be Cited?*." *Social Science Quarterly* 89:802-816.

Murray, Fiona. 2010. "The Oncomouse That Roared: Hybrid Exchange Strategies as a Source of Distinction at the Boundary of Overlapping Institutions1." *American Journal of Sociology* 116:341-388.

Najman, Jake M. and Belinda Hewitt. 2003. "The Validity of Publication and Citation Counts for Sociology and Other Selected Disciplines." *Journal of Sociology* 39:62-80.

National Academies of Science, National Academy of Engineering, and Institute of Medicine. 2005. *Facilitating Interdisciplinary Research*. Washington, DC: The National Academies Press.

National Research Council. 2014. *Convergence: Facilitating Transdisciplinary Integration of Life Sciences, Physical Sciences, Engineering, and Beyond* Washington, DC The National Academies Press.

Negro, Giacomo and Ming D. Leung. 2013. "Actual and Perceptual Effects of Category Spanning." *Organization Science* 24:684-696.

Negro, Giacomo, Michael T. Hannan, and Hayagreeva Rao. 2010. "Categorical contrast and audience appeal: Niche width and critical success in winemaking." *Industrial and Corporate Change* 19:1397–1425.

Owen-Smith, Jason and Walter W. Powell. 2003. "The Expanding Role of University Patenting in the life Sciences: Assessing the Importance of Experience and Connectivity." *Research Policy* 32:1695-1711.

Pelled, Lisa Hope, Kathleen M Eisenhardt, and Katherine R Xin. 1999. "Exploring the black box: An analysis of work group diversity, conflict and performance." *Administrative Science Quarterly* 44:1-28.

Phillips, Damon. 2011. "Jazz and the Disconnected: City Structural Disconnectedness and the Emergence of a Jazz Canon, 1897–1933." *American Journal of Sociology* 117:420-483.

Phillips, Damon and Ezra Zuckerman. 2001. "Middle Status Conformity: Theoretical Restatement and Evidence from Two Markets." *American Journal of Sociology* 107:379-429.

Poincare, Henri. 1952 [1908]. "Mathematical Creation." in *The Creative Process*, edited by B. Ghiselin. Berkeley, CA: University of California Press.

Ponomariov, Branco L. and P. Craig Boardman. 2010. "Influencing scientists' collaboration and productivity patterns through new institutions: University research centers and scientific and technical human capital." *Research Policy* 39:613-624.

Pontikes, Elizabeth G. 2012. "Two Sides of the Same Coin How Ambiguous Classification Affects Multiple Audiences' Evaluations." *Administrative Science Quarterly* 57:81-118.

Porter, Alan L., Alex S. Cohen, J. David Roessner, and Marty Perreault. 2007. "Measuring Research Interdisciplinarity." *Scientometrics* 72:117-147.

Porter, Alan L. and Ismael Rafols. 2009. "Is Science Becoming More Interdisciplinary? Measuring and Mapping Six Research Fields Over Time." *Scientometrics* 81:719-745.

Powell, Walter W., Kenneth W. Koput, and Laurel Smith-Doerr. 1996. "Interorganizational Collaboration and the Locus of Innovation: Networks of Learning in Biotechnology." *Administrative Science Quarterly* 41:116-145.

Pray, Leslie 2002. "Interdisciplinarity in Science and Engineering: Academia in Transition." *Science Career Magazine*.

Prpic, Katarina. 2002. "Gender and Productivity Differentials in Science." *Scientometrics* 55:27-58.

Rafols, Ismael, Loet Leydesdorff, Alice OHare, Paul Nightingale, and Andy Stirling. 2012. "How journal rankings can suppress interdisciplinary research: A comparison between Innovation Studies and Business & Management." *Research Policy* 41:1262-1282.

Rafols, Ismael and Martin Meyer. 2010. "Diversity and network coherence as indicators of interdisciplinarity: Case studies in bionanoscience." *Scientometrics* 82:263-287.

Rhoten, Diana and Andrew Parker. 2004. "Risks and rewards of an interdisciplinary research path." *Science(Washington)* 306:2046.

Rhoten, Diana and Stephanie Pfirman. 2007. "Women in interdisciplinary science: Exploring preferences and consequences." *Research Policy* 36:56-75.

Rinia, Ed, Thed van Leeuwen, and Anthony F van Raan. 2002. "Impact Measures of Interdisciplinary Research in Physics." *Scientometrics* 53:241-248.

Rosenkopf, Lori and Paul Almeida. 2003. "Overcoming Local Search Through Alliances and Mobility." *Management Science* 49:754-766.

Sá, Creso M. 2008. "'Interdisciplinary strategies' in US research universities." *Higher Education* 55:537-552.

Sanz, Luis Menendez , Maria  Bordons, and M. Angeles Zulueta. 2001. "Interdisciplinarity as a Multidimensional Concept: Its Measure in three Different Research Areas." *Research Evaluation* 10:47-58.

Sauer, Raymond D. 1988. "Estimates of the Returns to Quality and Coauthorship in Economic Academia." *Journal of Political Economy* 96:855-866.

Schilling, Melissa A. and Elad Green. 2011. "Recombinant Search and breakthrough Idea Generation: An Analysis of High Impact Papers in the Social Sciences." *Research Policy* 40:1321-1331.

Schumpeter, Joseph A. 1912 [1934]. *The Theory of Economic Development*. Cambridge, MA: Harvard Univesity Press.

Shi, Xiaolin, Lada A. Adamic, Belle L. Tseng, and Gavin S. Clarkson. 2009. "The Impact of Boundary Spanning Scholarly Publications and Patents." *PloS ONE* 4:e6547.

Shrum, Wesley, Joel Genuth, and Ivan Chompalov. 2007. *Structures of scientific collaboration*: MIT Press.

Singh, Jasjit and Lee Fleming. 2010. "Lone inventors as sources of breakthroughs: Myth or reality?" *Management Science* 56:41-56.

Trapido, Denis. in press. "When Audiences Value Novelty in Knowledge: Consistent Identities and Recognition Spillovers." *Research Policy*.

Uzzi, Brian, Satyam Mukherjee, Michael Stringer, and Ben Jones. 2013. "Atypical Combinations and Scientific Impact." *Science* 342:468-472.

van Dalen, Hendrik and Kene Henkens. 2005. "Signals in Science - On the Importance of Signaling in Gaining Attenion in Science." *Scientometrics* 64:209-233.

van Rijnsoever, Frank J and Laurens K Hessels. 2011. "Factors associated with disciplinary and interdisciplinary research collaboration." *Research Policy* 40:463-472.

Weick, Karl E. 1979. *The Social Psychology of Organizing*. Reading, MA: Addison-Wesley.

Weitzman, Martin L. 1998. "Recombinant Growth." *The Quarterly Journal of Economics* 113:331-360.

Wuchty, Stefan, Benjamin F. Jones, and Brian Uzzi. 2007. "The Increasing Dominance of Teams in Production of Knowledge." *Science* 316:1036-1039.

Zitt, Michel, Suzy Ramanana-Rahary, and Elise Bassecoulard. 2005. "Relativity of citation performance and excellence measures: From cross-field to cross-scale effects of field-normalisation." *Scientometrics* 63:373-401.

Zuckerman, Ezra W. 2005. "Typecasting and Generalism in Firm and Market: Genre-Based Career Concentration in the Feature Film Industry, 1933–1995." *Research in the Sociology of Organizations* 23:173-216.

Zuckerman, Ezra W. 1999. "The Categorical Imperative: Securities Analysts and the Illegitimacy Discount." *American Journal of Sociology* 104:1398-1438.

Zuckerman, Ezra W., Tai-Young Kim, Kalinda Ukanwa, and James von Rittmann. 2003. "Robust Identities or Nonentities?  Typecasting in the Feature-Film Labor Market." *American Journal of Sociology* 108:1018-1074.

Table 1 Panel A. Correlations and Descriptive Statistics.  N=854 scientists

| | Mean | S.D. | Visib. | Prod. | IDR | Field IDR | Mean JIF | Prof. Age | PhD rank | Status | Female | Field size | Field cites | Turnaround |
|---|---|---|---|---|---|---|---|---|---|---|---|---|---|---|
| Visibility (total # citations)[a] | 930 | 1779 | 1 | | | | | | | | | | | |
| Productivity (total # articles)[b] | 46 | 60 | 0.83 | 1 | | | | | | | | | | |
| Interdisciplinarity (mean IDR score)[b] | 0.63 | 0.10 | 0.05 | -0.05 | 1 | | | | | | | | | |
| Field Level IDR | 0.62 | 0.04 | -0.08 | -0.14 | 0.13 | 1 | | | | | | | | |
| Mean Journal Impact Factor (JIF) | 1.68 | 1.21 | 0.38 | 0.21 | 0.25 | 0.05 | 1 | | | | | | | |
| Professional Age (years since PhD) | 24 | 11 | 0.22 | 0.41 | -0.17 | -0.22 | -0.12 | 1 | | | | | | |
| Ranking of PhD-Granting Institution | 2.84 | 1.70 | 0.12 | 0.08 | 0.06 | 0.02 | 0.10 | 0.09 | 1 | | | | | |
| Status (N NAS members at center's university) | 34 | 39 | 0.08 | 0.05 | -0.02 | 0.01 | 0.05 | 0.05 | 0.35 | 1 | | | | |
| Female (=1, male=0) | 0.13 | 0.34 | -0.08 | -0.14 | 0.06 | 0.12 | 0.03 | -0.22 | 0.02 | 0.00 | 1 | | | |
| Field size N of PhDs) | 26,092 | 15,658 | 0.06 | 0.06 | -0.07 | 0.01 | 0.18 | 0.03 | 0.01 | -0.05 | 0.01 | 1 | | |
| Average Citations by Field | 11.52 | 6.17 | 0.13 | 0.11 | 0.20 | 0.12 | 0.32 | 0.06 | -0.01 | -0.02 | 0.10 | 0.20 | 1 | |
| Journal Turnaround Time by Field | 5.68 | 1.31 | -0.07 | -0.09 | 0.13 | 0.08 | -0.09 | -0.05 | 0.04 | -0.04 | 0.08 | -0.08 | -0.34 | 1 |

[a] as of 2010

[b] as of 2005

Table 1 Panel B. Correlations and Descriptive Statistics.  n=9647 person-years [c]

| | Mean | S.D. | Visib. | Prod. | IDR | N Authors | Mean JIF | Repeat Collab. | Reach | Lagged Visib. | Lagged Prod. | Prof Age | PhD rank | Status | Female | Field size | Field IDR | Turnaround | Field citn |
|---|---|---|---|---|---|---|---|---|---|---|---|---|---|---|---|---|---|---|---|
| Visibility (total # citations) | 81 | 180 | 1 | | | | | | | | | | | | | | | | |
| Productivity (total # articles) | 3 | 3 | 0.58 | 1 | | | | | | | | | | | | | | | |
| Interdisciplinarity (mean IDR score) | 0.62 | 0.15 | 0.06 | 0.02 | 1 | | | | | | | | | | | | | | |
| Number of Authors (avg per year) | 7.50 | 40.01 | 0.12 | 0.16 | 0.04 | 1 | | | | | | | | | | | | | |
| Mean Journal Impact Factor (JIF) | 1.69 | 1.80 | 0.31 | 0.14 | 0.14 | 0.19 | 1 | | | | | | | | | | | | |
| Repeat Collaborations (avg per year) | 0.15 | 0.28 | 0.01 | 0.07 | -0.04 | 0.10 | -0.06 | 1 | | | | | | | | | | | |
| Potential Reach (N focal paper SCs) | 1.60 | 0.75 | -0.07 | -0.07 | 0.23 | -0.05 | -0.11 | -0.02 | 1 | | | | | | | | | | |
| Visibility (cumulative, lagged) | 482 | 959 | 0.29 | 0.43 | 0.07 | 0.07 | 0.19 | -0.03 | -0.04 | 1 | | | | | | | | | |
| Productivity (cumulative, lagged) | 23.95 | 35.88 | 0.30 | 0.57 | 0.03 | 0.05 | 0.13 | -0.02 | -0.04 | 0.62 | 1 | | | | | | | | |
| Professional Age (years since PhD) | 13.02 | 10.39 | 0.05 | 0.17 | -0.03 | 0.04 | -0.03 | -0.01 | 0.01 | 0.47 | 0.48 | 1 | | | | | | | |
| Ranking of PhD-Granting Institution | 3.01 | 1.75 | 0.07 | 0.02 | 0.02 | -0.02 | 0.08 | -0.02 | 0.01 | 0.10 | 0.03 | 0.04 | 1 | | | | | | |
| Status (# NAS members at university in 2000) | 34.24 | 37.94 | 0.05 | 0.04 | -0.02 | -0.01 | 0.03 | -0.05 | 0.07 | 0.05 | 0.04 | -0.01 | 0.36 | 1 | | | | | |
| Female (=1, male=0) | 0.10 | 0.30 | -0.02 | -0.05 | 0.01 | -0.01 | 0.01 | -0.01 | -0.03 | -0.11 | -0.09 | -0.16 | -0.01 | 0.01 | 1 | | | | |
| Field size (N of PhDs) | 26632 | 15821 | 0.04 | -0.01 | -0.04 | -0.04 | 0.16 | -0.05 | -0.05 | 0.09 | 0.03 | 0.01 | 0.04 | -0.06 | 0.03 | 1 | | | |
| Field Level IDR | 0.62 | 0.04 | 0.03 | 0.01 | 0.28 | 0.00 | 0.07 | 0.01 | 0.06 | 0.03 | -0.01 | 0.00 | 0.03 | 0.02 | 0.10 | -0.18 | 1 | | |
| Journal Turnaround Time by Field | 6 | 1 | -0.04 | -0.07 | 0.08 | -0.01 | -0.03 | 0.02 | 0.00 | -0.07 | -0.08 | -0.05 | 0.01 | -0.02 | 0.08 | -0.03 | 0.28 | 1 | |
| Average Citations by Field | 11.67 | 6.10 | 0.10 | 0.09 | 0.16 | -0.02 | 0.20 | -0.05 | -0.03 | 0.14 | 0.10 | 0.03 | -0.01 | -0.06 | 0.08 | 0.18 | 0.47 | -0.34 | 1 |

[c] Correlations greater than .02 are significant at the .05 level

Table 2.  Path Analytic Models (Unstandardized Coefficients and Standard Errors)

| | Model 1 | Model 2 | Model 3 | Model 4 | Model 5 |
| --- | --- | --- | --- | --- | --- |
| | | | | Electrical Engineers | Life Scientists |
| | | SD of citations[b] | N of SCs[a] | | |
| **Effects on VISIBILITY (logged)** | | | | | |
| INTERDISCIPLINARY RESEARCH (IDR) | 0.69*** | 20.75*** | 0.01*** | 0.15 | 1.23 |
| | (0.26) | (6.74) | (0.00) | (0.37) | (0.99) |
| Productivity | 1.27*** | 6.45*** | 1.12*** | 1.34*** | 1.19*** |
| | (0.03) | (0.66) | (0.03) | (0.06) | (0.07) |
| Mean Journal Impact Factor (JIF) | 0.33*** | 7.42*** | 0.32*** | 0.40*** | 0.21*** |
| | (0.02) | (0.46) | (0.02) | (0.06) | (0.04) |
| Professional Age (years since PhD) | -0.01*** | -0.03 | -0.01*** | -0.02*** | -0.02** |
| | (0.00) | (0.07) | (0.00) | (0.00) | (0.01) |
| Ranking of PhD-Granting Institution | 0.02 | 0.48 | 0.01 | 0.03 | 0.04 |
| | (0.02) | (0.40) | (0.01) | (0.03) | (0.05) |
| Status (# NAS members at center's university) | 0.00** | 0.06*** | 0.00** | 0.00* | 0.00 |
| | (0.00) | (0.02) | (0.00) | (0.00) | (0.00) |
| Female (yes=1) | 0.09 | 0.85 | 0.07 | 0.11 | -0.04 |
| | (0.07) | (1.93) | (0.07) | (0.16) | (0.16) |
| Field size (number of PhDs) | -0.00*** | -0.00 | -0.00*** | -- | -- |
| | (0.00) | (0.00) | (0.00) | | |
| Field Level IDR | -1.35* | -59.05*** | -1.70** | -- | -- |
| | (0.81) | (21.24) | (0.78) | | |
| Field Citations | 0.01 | 0.28** | 0.01 | -- | -- |
| | (0.01) | (0.14) | (0.01) | | |
| Intercept | 1.79*** | 13.84 | 2.52*** | 0.91*** | 1.14 |
| | (0.47) | (12.36) | (0.48) | (0.28) | (0.73) |
| **Effects on PRODUCTIVITY (logged)** | | | | | |
| INTERDISCIPLINARY RESEARCH (IDR) | -0.77** | -0.62* | 0.04*** | -1.07** | -1.07 |
| | (0.35) | (0.35) | (0.00) | (0.52) | (1.67) |
| Mean Journal Impact Factor (JIF) | 0.22*** | 0.17*** | 0.03 | 0.15* | 0.02 |
| | (0.03) | (0.02) | (0.02) | (0.08) | (0.06) |
| Professional Age (years since PhD) | 0.05*** | 0.05*** | 0.04*** | 0.04*** | 0.04*** |
| | (0.00) | (0.00) | (0.00) | (0.01) | (0.01) |
| Ranking of PhD-Granting Institution | 0.07*** | 0.07*** | 0.02 | 0.07* | 0.10 |
| | (0.02) | (0.02) | (0.01) | (0.04) | (0.08) |
| Status (# NAS members at center's university) | -0.00 | -0.00 | -0.00 | -0.00 | -0.00 |
| | (0.00) | (0.00) | (0.00) | (0.00) | (0.00) |
| Female (yes=1) | -0.33*** | -0.34*** | -0.23*** | -0.11 | -0.50* |
| | (0.10) | (0.10) | (0.07) | (0.22) | (0.26) |
| Field size (number of PhDs) | 0.00 | 0.00* | 0.00 | -- | -- |
| | (0.00) | (0.00) | (0.00) | | |
| Field Level IDR | -0.74 | -0.40 | -4.41*** | -- | -- |
| | (0.96) | (0.95) | (0.66) | | |
| Field level Journal Turnaround Time | -0.02 | -0.03 | -0.00 | -- | -- |
| | (0.03) | (0.03) | (0.02) | | |
| Intercept | 2.53*** | 2.44*** | 3.72*** | 2.44*** | 2.93** |
| | (0.55) | (0.55) | (0.40) | (0.34) | (1.20) |
| **Effects on INTERDISCIPLINARY RESEARCH** | | | | | |
| Mean Journal Impact Factor (JIF) | 0.01*** | 0.01*** | 4.87*** | 0.05*** | -0.00 |
| | (0.00) | (0.00) | (0.51) | (0.01) | (0.00) |
| Professional Age (years since PhD) | -0.00*** | -0.00*** | 0.39*** | -0.00 | 0.00 |
| | (0.00) | (0.00) | (0.06) | (0.00) | (0.00) |
| Ranking of PhD-Granting Institution | 0.00 | 0.00 | 1.37*** | 0.00 | 0.01 |
| | (0.00) | (0.00) | (0.38) | (0.01) | (0.01) |
| Status (# NAS members at center's university) | -0.00*** | -0.00*** | 0.02 | -0.00*** | -0.01*** |
| | (0.00) | (0.00) | (0.05) | (0.00) | (0.00) |
| Status$^2$ | 0.00*** | 0.00*** | -0.00 | 0.00*** | 0.00*** |
| | (0.00) | (0.00) | (0.00) | (0.00) | (0.00) |
| Female (yes=1) | -0.00 | -0.00 | -2.72 | -0.03 | -0.00 |
| | (0.01) | (0.01) | (1.84) | (0.03) | (0.02) |
| Field IDR | 0.91*** | 0.91*** | 77.27*** | -- | -- |
| | (0.08) | (0.08) | (16.16) | | |
| Intercept | 0.09* | 0.09* | -34.75*** | 0.55*** | 0.73*** |
| | (0.05) | (0.05) | (10.11) | (0.03) | (0.03) |
| N Individuals | 854 | 854 | 854 | 160 | 78 |
| Chi-square test statistic (T) | 15.9 | 49.1 | 17.3 | 4.2 | 4.9 |
| Degrees of freedom | 78 | 87 | 75 | 34 | 37 |
| AIC | 57289 | 65110 | 65684 | 6803 | 3018 |
| BIC | 57749 | 65628 | 66145 | 6963 | 3140 |
| CFI (ideal=1) | 1.0 | 1.0 | 1.0 | 1.0 | 1.0 |
| 1-RMSEA (ideal=1) | 1.0 | 1.0 | 1.0 | 1.0 | 1.0 |

* p<0.10  ** p<.05  *** p<.01 (two-tailed tests)

[a] For this model we use the number of Subject Categories instead of IDR as the IV.

[b] For this model we use the standard deviation of journal impact factor rather than its mean in the visibility equation.

Table 3: Regressions of Productivity and Visibility on IDR[a]

| | Model 1:[b] FE Productivity | Model 2:[c] PA Productivity | Model 3:[c] PA Productivity | Model 4:[d] FE Visibility | Model 5:[e] RE Visibility | Model 6:[e] RE Visibility |
|---|---|---|---|---|---|---|
| IDR | -0.143** | -0.146** | -0.129* | 0.468*** | 0.254** | 0.302*** |
| | (0.071) | (0.070) | (0.069) | (0.063) | (0.122) | (0.121) |
| Productivity (cumulative, lagged) | -0.001*** | - | - | 0.000 | 0.007*** | 0.007*** |
| | (0.000) | - | - | (0.000) | (0.001) | (0.001) |
| Visibility (cumulative, logged and lagged) | 0.108*** | 0.133*** | 0.133*** | -0.052*** | 0.228*** | 0.228*** |
| | (0.012) | (0.011) | (0.011) | (0.011) | (0.012) | (0.012) |
| Number of authors | 0.001*** | 0.002*** | 0.002*** | 0.000 | 0.002*** | 0.002*** |
| | (0.000) | (0.000) | (0.000) | (0.000) | (0.001) | (0.001) |
| Journal Impact Factor | -0.004 | 0.001 | 0.001 | 0.158*** | 0.212*** | 0.213*** |
| | (0.005) | (0.005) | (0.005) | (0.009) | (0.017) | (0.017) |
| Repeat Collaborations | 0.296*** | 0.331*** | 0.332*** | -0.008 | 0.303*** | 0.305*** |
| | (0.027) | (0.028) | (0.028) | (0.025) | (0.056) | (0.056) |
| Potential Reach (N focal paper SCs) | | | | 0.035** | -0.004 | -0.004 |
| | | | | (0.015) | (0.021) | (0.020) |
| Female | | -0.088 | -0.091 | | -0.001 | -0.005 |
| | | (0.057) | (0.057) | | (0.066) | (0.066) |
| Professional Age | | -0.003 | -0.003 | | -0.022*** | -0.022*** |
| | | (0.002) | (0.002) | | (0.003) | (0.003) |
| Ranking of PhD institution | | 0.022 | 0.021 | | 0.029** | 0.029** |
| | | (0.014) | (0.014) | | (0.014) | (0.014) |
| Status of Center institution | | 0.000 | 0.000 | | 0.001 | 0.001 |
| | | (0.001) | (0.001) | | (0.001) | (0.001) |
| Field Size | | 0.000 | 0.000 | | 0.000 | 0.000 |
| | | (0.000) | (0.000) | | (0.000) | (0.000) |
| Field Level IDR | | 0.858* | 0.877* | | 0.968 | 1.075 |
| | | (0.475) | (0.475) | | (0.694) | (0.697) |
| IDR * Field Level IDR | | | 3.428** | | | 7.286*** |
| | | | (1.569) | | | (2.605) |
| Journal Turnaround Time by Field | | -0.036** | -0.036** | | | |
| | | (0.016) | (0.016) | | | |
| Average citations by field | | | | | 0.011*** | 0.010** |
| | | | | | (0.004) | (0.005) |
| Year 1985-89 | -0.046 | -0.133*** | -0.133*** | | | |
| | (0.039) | (0.041) | (0.041) | | | |
| Year 1990-94 | 0.080* | -0.047 | -0.047 | | | |
| | (0.042) | (0.050) | (0.050) | | | |
| Year 1995-99 | 0.195*** | 0.045 | 0.044 | | | |
| | (0.046) | (0.052) | (0.052) | | | |
| Year 2000-04 | 0.296*** | 0.111** | 0.109* | | | |
| | (0.052) | (0.057) | (0.058) | | | |
| Intercept | 1.730*** | 0.008 | 0.447*** | 1.700*** | 0.968*** | 1.508*** |
| | (0.082) | (0.306) | (0.111) | (0.113) | (0.511) | (0.517) |
| N Obs | 8789 | 9647 | 9647 | 29782 | 9647 | 9647 |
| N Individuals | 804 | 854 | 854 | 854 | 854 | 854 |
| F | 75.11 | 30.77 | 29.82 | 34.20 | 49.64 | 49.61 |

* p<0.10  ** p<.05  *** p<.01 (two-tailed tests)

[a] All models include multiple imputation and exclude observations with IDR=0.

[b] Model 1 is Negative Binomial Regression at the person-year level with person-fixed effects.

[c] Models 2-3 are Population-Averaged Negative Binomial Regressions at the person-year level with robust standard errors. Models do not converge with lagged cumulative productivity.

[d] Model 4 is GLS regression at the paper-level with person and year fixed effects and robust standard errors.

[e] Models 5-6 are Random Effects GLS Regressions at the person-year level with year fixed effects and robust standard errors.

## Appendix A1. Construction of the Interdisciplinary Research (IDR) Measure for 3 Hypothetical Articles

**IDR = 1 - ∑ij sij pi pj**

Porter's measure of Integration, above, is used to capture the extent to which a paper is interdisciplinary. This measure incorporates the variety (i.e., number) of disciplines, their balance (i.e., the evenness of the distribution), and -- uniquely -- their similarity (i.e., their cognitive distance) into one index. Disciplines are proxied by the 244 WoS Subject Categories (SCs; examples include Sociology, Management, and Chemical Engineering). In this example we assume there are only 4 WoS SCs, and that each paper has only 5 references. As input, we pool all SCs from the focal paper's set of references to get variety (captured by the number of SCs referenced) and balance (captured by Pi, the proportion of references falling in SCi). We then incorporate the similarity scores for each SC-SC combination (Sij, from a SCxSC co-citation matrix which we convert to cosines and normalize) which are derived from the population of all WoS-indexed articles and thus shared by all focal papers; we highlight it in grey below. A paper's IDR score increases as it references more, relatively unrelated SCs (Porter, Cohen, Roessner, and Perreault 2007:277). The IDR score ranges from 0 to 1, with scores closer to 1 indicating greater interdisciplinarity.

### Article 1.

This paper references all 4 SCs, two of which are distant (SC2 and SC4, Sij=0.0001), and thus has a high IDR score:

| | SC1 | SC2 | SC3 | SC4 |
|---|---|---|---|---|
| Ref 1 | | x | | |
| Ref 2 | x | | | |
| Ref 3 | | | x | |
| Ref 4 | | | | x |
| Ref 5 | x | | | |
| **Pi =** | 0.4 (=2/5) | 0.2 | 0.2 | 0.2 |
| **Pi*Pj =** | | | | |
| SC1 | 0.16 (=0.4*0.2) | | | |
| SC2 | 0.08 | 0.04 | | |
| SC3 | 0.08 | 0.04 | 0.04 | |
| SC4 | 0.08 | 0.04 | 0.04 | 0.04 |
| **Sij =** | | | | |
| SC1 | 1 | | | |
| SC2 | 0.2487 | 1 | | |
| SC3 | 0.0083 | 0.3503 | 1 | |
| SC4 | 0 | 0.0001 | 0.0011 | 1 |
| **Sij * Pi * Pj =** | | | | |
| SC1 | 0.16 | | | |
| SC2 | 0.0199 | 0.04 | | |
| SC3 | 0.0007 | 0.0140 | 0.04 | |
| SC4 | 0.0000 | 0.0000 | 0.0000 | 0.04 |
| **∑ij Sij * Pi * Pj =** | 0.3146 | | | |
| **1 - ∑ij Sij * Pi * Pj =** | **0.6854** | | | |

### Article 2.

This paper references 2 highly related SCs, (SC2 and SC3, Sij=0.35, which is one of the highest cosines in our data), and thus has a low IDR score:

| | SC1 | SC2 | SC3 | SC4 |
|---|---|---|---|---|
| Ref 1 | | x | | |
| Ref 2 | | | x | |
| Ref 3 | | | x | |
| Ref 4 | | | x | |
| Ref 5 | | | x | |
| **Pi =** | 0 | 0.2 | 0.8 | 0 |
| **Pi*Pj =** | | | | |
| SC1 | 0 | | | |
| SC2 | 0 | 0.04 | | |
| SC3 | 0 | 0.16 | 0.64 | |
| SC4 | 0 | 0 | 0 | 0 |
| **Sij =** | | | | |
| SC1 | 1 | | | |
| SC2 | 0.2487 | 1 | | |
| SC3 | 0.0083 | 0.3503 | 1 | |
| SC4 | 0 | 0.0001 | 0.0011 | 1 |
| **Sij * Pi * Pj =** | | | | |
| SC1 | 0.00 | | | |
| SC2 | 0.00 | 0.04 | | |
| SC3 | 0.0000 | 0.0561 | 0.64 | |
| SC4 | 0.0000 | 0.0000 | 0.0000 | 0.00 |
| **∑ij Sij * Pi * Pj =** | 0.7361 | | | |
| **1 - ∑ij Sij * Pi * Pj =** | **0.2639** | | | |

### Article 3.

This paper references 2 distant SCs (SC3 and SC4, Sij=0.0011), so its IDR score is higher than Article 2's IDR score:

| | SC1 | SC2 | SC3 | SC4 |
|---|---|---|---|---|
| Ref 1 | | | | x |
| Ref 2 | | | x | |
| Ref 3 | | | x | |
| Ref 4 | | | x | |
| Ref 5 | | | x | |
| **Pi =** | 0 | 0 | 0.8 | 0.2 |
| **Pi*Pj =** | | | | |
| SC1 | 0 | | | |
| SC2 | 0 | 0 | | |
| SC3 | 0 | 0 | 0.64 | |
| SC4 | 0 | 0 | 0.16 | 0.04 |
| **Sij =** | | | | |
| SC1 | 1 | | | |
| SC2 | 0.2487 | 1 | | |
| SC3 | 0.0083 | 0.3503 | 1 | |
| SC4 | 0 | 0.0001 | 0.0011 | 1 |
| **Sij * Pi * Pj =** | | | | |
| SC1 | 0.00 | | | |
| SC2 | 0.0000 | 0.00 | | |
| SC3 | 0.0000 | 0.0000 | 0.64 | |
| SC4 | 0.0000 | 0.0000 | 0.0002 | 0.04 |
| **∑ij Sij * Pi * Pj =** | 0.6802 | | | |
| **1 - ∑ij Sij * Pi * Pj =** | **0.3198** | | | |

Appendix A2. Comparing our sample to broader populations of scientists & scientific papers (t-tests)

| | Our sample | | SESTAT scientists[a] |
|---|---|---|---|
| Gender | 0.13 | *** | 0.034 |
| Professional Age | 18.6 | *** | 15.3 |
| Productivity | 13.1 | *** | 7.9 |
| N individuals | 761[b] | | 5,520 |

| | | Our sample | Web of Science papers[c] |
|---|---|---|---|
| Uzzi's Conventionality (median value of the median $z$-score) | 1980s | 68.6 [d] | 69 |
| | 1990s | 98.6 | 99.5 |
| Uzzi's Novelty (% of papers with a 10th percentile $z$ score <0) | 1980s | 29 | 40.8 |
| | 1990s | 34.2 | 40.7 |
| N publications | | 19,823 | |

[a] Restricted to the years (1998-2003) available for SESTAT.  Our productivity data come from Web of Science; SESTAT productivity data come from survey self-reports.
[b] Varies slightly depending on variable, due to missing data.
[c] Based on all articles published in WOS as reported in Uzzi et al. (2013) by decade.  Our sample is restricted to relevant decade for comparison.
[d] We do not have the standard deviation of the population of papers reported in Uzzi et al. (2013), so we cannot conduct a t-test.

Appendix A3. Robustness Checks: Regressions of Productivity and Visibility on IDR[a]

|  | Model 1: FE Visibility | Model 2: FE Productivity | Model 3: FE Visibility |
|---|---|---|---|
| IDR | 0.276** | -0.312* | 0.485*** |
|  | (0.130) | (0.160) | (0.070) |
| Pre/Post Center Formation (Post=1) | -0.221** | 0.121 |  |
|  | (0.110) | (0.120) |  |
| Post-Center*IDR | 0.257 | 0.119 |  |
|  | (0.160) | (0.180) |  |
| Conventionality+ Novelty (Uzzi measure) |  |  | 0.097*** |
|  |  |  | (0.020) |
| Cumulative Publications (lagged) | -0.005** | -0.003*** | 0.000 |
|  | (0.001) | (0.000) | (0.000) |
| Cumulative Citations (lagged) | -0.040*** | 0.070*** | -0.051*** |
|  | (0.002) | (0.020) | (0.001) |
| Mean Number of authors | 0.000 | 0.001*** | 0.000 |
|  | 0.000 | (0.000) | (0.000) |
| Mean Journal Impact Factor | 0.138*** | -0.004 | 0.156*** |
|  | (0.010) | (0.010) | (0.010) |
| Proportion of Repeat Collaboration | 0.054 | 0.194*** | 0.007 |
|  | (0.040) | (0.050) | (0.020) |
| Number of Subject Categories | 0.018 | -0.021 | 0.019 |
|  | (0.020) | (0.020) | (0.020) |
| Intercept | 2.519*** | 2.653*** | 1.911*** |
|  | (0.120) | (0.180) | (0.270) |
| N obs | 10468 | 3269 | 29102 |
| N Indiv | 689 | 573 | 854 |
| F | 26.08 | 118.52 | 34.51 |

[a] Visibility is modeled at the paper level and productivity at the person-year level with person fixed-effects. Publications included 5 years prior and post center formation.

*** $p<.01$; ** $p<.05$; * $p<.10$ (two-tailed tests).